\documentclass[prc,preprint,showpacs,showkeys,superscriptaddress,nofootinbib,tightenlines]{revtex4}

\usepackage{graphicx,color}  %
\usepackage{bm}  %                                          

\newcommand{\beq}{\begin{equation}}
\newcommand{\eeq}{\end{equation}}
\newcommand{\bea}{\begin{eqnarray}}
\newcommand{\eea}{\end{eqnarray}}
\newcommand{\ben}{\begin{eqnarray*}}
\newcommand{\een}{\end{eqnarray*}}

\newcommand{\simge}{\hspace*{0.2em}\raisebox{0.5ex}{$>$}
     \hspace{-0.8em}\raisebox{-0.3em}{$\sim$}\hspace*{0.2em}}
\newcommand{\simle}{\hspace*{0.2em}\raisebox{0.5ex}{$<$}
     \hspace{-0.8em}\raisebox{-0.3em}{$\sim$}\hspace*{0.2em}}

\begin{document}

\title{Two and Three Nucleons in a Trap and the Continuum Limit}

\author{J. Rotureau}
\affiliation{Department of Physics, University of Arizona, 
Tucson, AZ 85721, USA}

\author{I. Stetcu}
\affiliation{Department of Physics, University of Washington, Box 351560, 
Seattle, WA 98195-1560, USA}
\affiliation{Theoretical Division, Los Alamos National Laboratory, 
Los Alamos, New Mexico 87545, USA}

\author{B.R. Barrett}
\affiliation{Department of Physics, University of Arizona, 
Tucson, AZ 85721, USA}

\author{U. van Kolck}
\affiliation{Department of Physics, University of Arizona, 
Tucson, AZ 85721, USA}

\begin{abstract}
We describe systems of two and three nucleons trapped in a harmonic-oscillator 
potential with interactions 
from the 
pionless effective field theory up to next-to-leading order (NLO).
We construct the two-nucleon interaction using two-nucleon scattering 
information.
We calculate the trapped levels in the three-nucleon system
with isospin $T=1/2$ 
and determine 
the three-nucleon force needed for stability of
the triton.
We extract neutron-deuteron phase shifts, and show that
the quartet scattering length 
is in good agreement 
with experimental data.
\end{abstract}

%\smallskip
%\pacs{03.75.Ss, 34.20.Cf, 21.60.Cs}
%\keywords{Few-nucleon systems, effective field theory}
\maketitle  

\section{Introduction}
\label{intro}

A major goal of nuclear physics is to derive nuclear structure 
``{\it ab initio}'',
that is, starting from inter-nucleon interactions consistent with QCD.
This requires a many-body technique that provides a numerical solution to
the Schr\"odinger equation for a system of $A$ interacting particles 
within a restricted space, which is sufficiently small
to be handled by accessible computers.
The no-core shell model (NCSM) \cite{NCSM}
is such a many-body technique, where
the restricted space is generated by harmonic-oscillator (HO) wavefunctions. 
In the traditional NCSM, effective inter-nucleon interactions 
adapted to the restricted
space are derived from a given potential.

The inter-nucleon potential is not directly observable; it is merely
an intermediate step to obtain measurable quantities.
Its very definition requires the choice of a restricted space,
and care is needed to make sure that measurable quantities are
independent of this choice, that is, are renormalization-group (RG) invariant.
The framework to accomplish this is effective field theory (EFT).
A potential 
constructed in EFT is improvable in a systematic expansion,
and can be used as input in many-body problems.
Some of current {\it ab initio} calculations do, indeed, start 
with a potential inspired by EFT, but they suffer from limited
(or no) RG invariance.
Alternatively,
we can construct manifestly RG-invariant observables
starting from inter-particle interactions defined directly within the 
restricted space of the 
many-body technique using the general principles of EFTs 
\cite {LO_eft_nuc,us3}.  
In Ref. \cite {LO_eft_nuc}, we have demonstrated this idea for
a NCSM-type restricted space
in leading order (LO) of the so-called pionless EFT.

At low energies, the physics of two- and few-body systems is insensitive to 
the details of the interaction at short distances. Thus, in the case of an 
interaction of finite range $R$, short-range details are irrelevant for 
the description of processes involving momenta $k\simle 1/R$. 
The pionless, or contact, EFT \cite{ARNPSreview}
uses this separation of scales to 
construct the
potential as a sum of delta functions
and their derivatives, which for observables translates into expansions 
in powers of $kR$. 
A particularly interesting class of systems is that where
the two-body $S$-wave scattering length $a_2$ is large, $a_2 \gg R$,
because then the 
LO potential solved exactly produces a 
real or virtual bound state at $k\simeq i/a_2 $. 
This EFT has been applied
to nuclear physics \cite{aleph,d_n_eft,pionlessT,pionless4}, 
where $a_2 \gg 1/m_\pi$, the inverse of the pion mass.
At larger momentum, $k\sim m_\pi$, pions need to be taken into account,
and the more sophisticated pionful, or chiral, EFT \cite{ARNPSreview}
is needed where, in addition to delta functions, pion exchange
is explicitly included.

The definition of the delta functions themselves is tied to the 
restricted space.
Their strengths depend on the 
size of the space.
Fitting the strengths of the two- and three-nucleon contact interactions 
that appear at LO in the pionless EFT
to reproduce the deuteron, triton and $^4$He ground-state binding energies, 
the energies of other
$^4{\rm{He}}$ and $^6{\rm{Li}}$ states were found in Ref. \cite {LO_eft_nuc}
to agree with experiment within the expected errors for a LO calculation 
($\sim 30\%$).

In order to demonstrate systematic improvement, one needs to calculate 
corrections beyond LO. However, beyond LO
the number of couplings in the EFT expansion increases significantly, 
making the fit to few-body 
binding energies impractical. 
Hence, we have developed an approach that requires only information 
from the two-nucleon system in order to fix the two-nucleon interactions. 
This can be done by considering the two-body system in a HO potential
and relating the energy levels to the scattering parameters.
In Refs. \cite{trap0,us1} we have constructed the two-body interaction up to  
next-to-next-to-leading order (NNLO),
which reproduces the lowest energy levels obtained from given scattering
parameters. 
We have used it to calculate the 
spectra of trapped systems of a few 
two-state fermions \cite{trap0,us2},
while systems of bosons were addressed in Ref. \cite{hanstrap}.

We have also 
suggested \cite{us3} that the same method can be used for nucleons.
Here we implement this approach. 
In fact, 
one of the goals of this paper is to show that meaningful results can be 
obtained for nuclear physics by trapping the nuclear system. 
This is not surprising, since in the middle of the trap the wavefunction 
has no knowledge of the trap's existence \cite{us1}. 
However,
while in atomic systems the trap is physical and dominates the long-range 
behavior, for nuclear systems the trap is just an artifact introduced 
in order to define 
the interaction. 
We therefore need an extra step at the end, that of making the trap
large, in order to extrapolate energies to the ``continuum'' limit.
The discrete states in the trap approach the untrapped
spectrum of a few bound states and a continuum of scattering states.  
Following the procedure devised in Ref. \cite{us2} for the atom-dimer
system, we extract 
the neutron-deuteron scattering length from the trapped-system levels
at NLO.  
This extends to the three-nucleon system the ``inverse'' connection
between trapped levels and scattering stressed at the two-body level
in Refs. \cite{us1,luu}.

A significant complication of the nuclear case, compared to the 
two-state-fermion case, is the role of three-body forces.
In the absence of the trap, a three-nucleon force is needed 
in the pionless EFT at LO in order to achieve RG invariance
\cite{pionlessT}. 
The same is true for bosons \cite{3bozos}, and it has been
shown in Ref. \cite{hanstrap} that this feature is not affected
by the presence of the trap, as expected from
the short-distance character of renormalization.
We show here that
the same holds for nucleons in the presence of the trap, and determine
the three-nucleon force needed for cutoff independence up to NLO.

The paper is organized as follows. 
After we set up our framework in  Sec. \ref{sec_gen},
we construct in Sec. \ref{sec_d} the two-nucleon interaction up to NLO
using the two-nucleon scattering data as input.
In Sec. \ref{sec_3n} we apply the formalism to 
three nucleons in two different channels described by
total isospin $T=1/2$ and total angular momentum/parity: 
{\it (i)} $J^{\pi}=3/2^{+}$, which is similar to the system of 
three two-component fermions and does not involve a three-body force 
up to NLO; and 
{\it (ii)} $J^\pi=1/2^+$, 
which requires a contact three-body force already in LO.
In the first channel, we calculate the quartet scattering length 
for deuteron-nucleon scattering, obtaining the same accuracy as 
similar continuum calculations, 
while for the second channel we demonstrate the collapse of the system, 
as in free space. 
We summarize and conclude in Sec. \ref{sec_last}.

\section{Preliminaries}
\label{sec_gen}

We consider a non-relativistic system of $A$ nucleons 
of mass $m_N$ trapped in a 
HO potential of frequency $\omega$,
or alternatively
of length 
\begin{equation}
b=\sqrt{\frac{2}{m_N \omega}}.
\end{equation}
The HO potential can be decomposed into two pieces, 
one acting on the center of mass (CM) of the particles 
and one on their relative coordinates. 
We denote by $\vec{r}_i$ ($\vec{p}_i$) the position (momentum) of particle 
$i$ with respect to the origin of the HO potential.
The Hamiltonian describing the relative motion of the particles 
is given by
\begin{eqnarray}
H= H_0+\sum_{i<j}V_{ij}+ \sum_{i<j<k}V_{ijk}+ \ldots  ,
\label{hami}
\end{eqnarray}
with
\begin{eqnarray}
H_0= \frac{\omega}{2A} \sum_{i<j} 
\left[\frac{b^2}{2}(\vec{p}_i -\vec{p}_j)^2 
+\frac{2}{b^2} (\vec{r}_i -\vec{r}_j)^2 \right].
\label{hami0}
\end{eqnarray}
Here $V_{ij}$ and $V_{ijk}$ denote the two-nucleon and three-nucleon
potentials, respectively, 
more-body interactions being lumped into ``$\ldots$''.

In pionless EFT, the inter-nucleon potential is expanded in derivatives
of delta functions, or powers of momenta in momentum space.
An important task in EFT is to provide a power counting for observables.
One can show \cite{aleph,d_n_eft,pionlessT,pionless4}
that in pionless EFT observables can be written as
expansions in $Q/M_{hi}$, where $Q$ denotes the generic
external momenta of the process under consideration
and $M_{hi}\sim m_\pi$ is the scale at which pion effects
need to be accounted for explicitly.
At any order of truncation, errors scale as $Q/M_{hi}$ to the 
power of the most important neglected order.
Of course, as in any theory one has to 
carry out renormalization.
Delta functions are singular and require
an ultraviolet (UV) momentum cutoff $\Lambda_A$ 
in order for them to be well defined.
Observables are (nearly) independent of the arbitrary cutoff as long as
the coefficients of the delta functions depend on the UV cutoff
appropriately ---and
as long as there are enough interactions at the given order,
which is a non-trivial consistency check on the power counting.
The truncation generates an error due to the cutoff that grows as 
$Q/\Lambda_A$,
so cutoff errors are minimized as $\Lambda_A$ increases.
However, one should keep in mind that there are always errors 
that grow as  $Q/M_{hi}$, which cannot be minimized by increasing the cutoff.

For convenience we diagonalize $H$ using HO wavefunctions.
We can work with Jacobi coordinates 
defined in terms of differences between the CM positions of 
sub-clusters within the $A$-body system, e.g.,
\begin{eqnarray}
\vec{\xi}_1&=&\sqrt{\frac{1}{2}}\left(\vec{r}_1-\vec{r}_2\right),
\nonumber\\
\vec{\xi}_2&=&\sqrt{\frac{2}{3}}
\left[\frac{1}{2}\left(\vec{r}_1+\vec{r}_2\right)-\vec{r}_3\right],
\nonumber\\
\vdots && 
\nonumber\\
\vec{\xi}_{A-1}&=& \sqrt{\frac{A-1}{A}}
\left[\frac{1}{A-1}\left(\vec{r}_1+\vec{r}_2+\cdots+\vec{r}_{A-1}\right)
-\vec{r}_{A}\right] .
\end{eqnarray}
In terms of them, the HO Hamiltonian (\ref{hami0}) becomes a collection
of $A-1$ HOs,
\begin{eqnarray}
H_0&=&
\frac{\omega}{2}
\sum_{\rho=1}^{A-1} 
\left[\left( \frac{b \vec{p}_{\xi_{\rho}}}{\sqrt{2}}\right)^{2}
+\left(\frac{\sqrt{2}\vec{\xi}_{\rho}}{b}\right)^{2} \right],
\end{eqnarray}
where $\vec{p}_{\xi_{\rho}}$ is the momentum canonically conjugated to 
$\vec{\xi}_{\rho}$.
We use
a basis made out
of properly antisymmetrized combinations of 
$A-1$ eigenfunctions $\phi_{n_i l_i}$
of $H_0$, which are characterized by the radial quantum numbers $n_i$
and the angular momenta $l_i$. The energy of a basis state can be written as 
$[N_A+3(A-1)/2]\omega$,
with $N_A$ an integer,
\begin{eqnarray}
N_A=\sum_{i=1}^{A-1}\left(2n_i+l_i\right).
\label{quanta_tot}
\end{eqnarray}
A numerical calculation can only be carried out with a finite
number of basis elements, which span the ``model space''.
We include in the model space all states up to a maximum
integer $N_A^{max}$, 
which provides a natural momentum cutoff \cite{LO_eft_nuc} 
\begin{equation}
\Lambda_A=\frac{1}{b} \sqrt{2N_A^{max}+3(A-1)}.
\label{LambdaA}
\end{equation} 
Since there is a minimum step in energy supplied by $\omega$, one can think
of the model space as containing also an infrared (IR) cutoff \cite{LO_eft_nuc}
\begin{equation}
\lambda=\frac{1}{b}.
\label{lambda}
\end{equation} 
Within the low energies where the EFT applies, the errors introduced
by the limited size of the model space should decrease as $\lambda$ decreases,
as well as $\Lambda_A$ increases.
This is the simple requirement that the HO oscillator be wide enough to
accommodate the nuclear states we are interested in.

As the trap is made larger, the states 
after diagonalization will approach their untrapped counterparts.
The lowest states become the free-nucleus bound states,
while states higher up in energy coalesce into a continuum of
scattering states.
Conversely, 
sufficiently near the center of a given trap wavefunctions
resemble those of the untrapped system. Yet, they must also
depend on the corresponding energy.
Thus, there is a connection between the energy levels in the trap
and the parameters that characterize the untrapped wavefunctions.
The latter are related to the the scattering, or $T$, matrix,
and thus to phase shifts, which 
at sufficiently low energies can be described
by effective-range expansion (ERE) parameters \cite{ERE}.
Thus, there is a relation between trap levels and phase shifts,
which we can explore in two ways.
First, we can use the scattering data as input to determine 
the levels inside the trap, and use a subset of the latter
to fix the coefficients of the inter-nucleon interactions \cite{trap0,us1}.
Other levels, at the same $A$ or not, can then be predicted.
Second, the predicted levels can be used to calculate
scattering phase shifts \cite{us1,us2}. 

In the remainder of the paper we carry out this method explicitly
to NLO in the pionless EFT for $A=2,3$.
We use the known two-nucleon ERE parameters to determine the 
two-nucleon interaction, and then extract neutron-deuteron 
scattering information.
In this first approach we do not include electromagnetic interactions
nor isospin-breaking effects, which are of higher order.
The same method to deal with nuclear systems can 
in the future be carried out
with the pionful EFT \cite{ARNPSreview}, where $M_{hi}$ is higher
and, consequently, denser systems can be handled.

\section{Two nucleons in a harmonic trap}
\label{sec_d}

We construct a two-body potential based on the ideas of EFT to describe the 
interaction between the two nucleons. 
The method is described in more detail for a single channel in 
Ref. \cite{us1}.

The two-nucleon interaction can be expanded as 
\begin{equation}
V(\vec{\xi}_1)
=\frac{C_0 }{2\sqrt{2}}\delta(\vec{\xi}_1)
-\frac{C_2}{4\sqrt{2}} 
\left[\left(\nabla^2\delta(\vec{\xi}_1)\right)
+\delta(\vec{\xi}_1)\nabla^2\right]
+\ldots ,
\label{V_two_body_eft}
\end{equation}
where $C_0$ and $C_2$ are parameters, and ``\ldots'' denote terms of 
higher orders. 
At LO, the Schr\"odinger equation of the trapped two-nucleon 
system is solved exactly with the potential given by the first term in 
the expression (\ref{V_two_body_eft}).
For a given ultraviolet cutoff $\Lambda_2$, the low-energy constant 
$C_0(\Lambda_2)$  is adjusted
such that one energy level of the two-nucleon system, 
which we take to be the lowest, is reproduced for all values of
$\Lambda_2$.
Corrections beyond leading orders should be calculated in increasing orders 
in perturbation theory. 
At NLO, in particular, corrections are obtained by considering the 
second term in Eq. (\ref{V_two_body_eft}) in first-order perturbation theory,
with $C_0(\Lambda_2)$ and $C_2(\Lambda_2)$ 
adjusted to reproduce two two-nucleon levels, here the lowest two.
Because the second term contributes to the ground state, 
the $\Lambda_2$ dependence of $C_0$ changes.
It is convenient to write 
$C_0(\Lambda_2)=C_0^{(0)}(\Lambda_2)+C_0^{(1)}(\Lambda_2)+\ldots$
and $C_2(\Lambda_2)=C_2^{(1)}(\Lambda_2)+\ldots$,
where the superscript $^{(n)}$ corresponds to the $\Lambda_2$ dependence
fixed at N$^{(n)}$LO. Only the $C_0^{(0)}(\Lambda_2)$ 
piece of the interaction
is iterated to all orders.

The interactions up to NLO affect only $S$ waves, higher waves coming at
NNLO and beyond.
Differently from the case of two-component fermions,
in the case of two nucleons there are two $L=0$ channels to 
consider, where the nucleons couple to total spin $S=0,1$. 
For a relative momentum $k \ll m_\pi$
the interaction in free space, {\it{i.e.}} when there is no trap,
gives rise \cite{aleph} to a phase shift $\delta_2(k)$ 
given by the ERE \cite{ERE},
\begin{equation}
k \, \cot \delta_2(k) =-\frac{1}{a_2}+\frac{r_2 }{2}k^{2}+ \ldots,
\label{ERE}
\end{equation}
where 
$a_2$, $r_2$, $\ldots$ are, respectively,
the scattering length, effective range, and higher ERE parameters 
not shown explicitly. 
The ERE parameters can be directly related \cite{aleph}
to the parameters in Eq. (\ref{V_two_body_eft}).
At LO, we obtain only the scattering-length term, while at NLO
the effective-range term appears as well.
In the following, we use the empirical values
$a_{2t}=5.425$ fm  and $r_{2t}=1.749$ fm  in the triplet
channel, and $a_{2s}=-18.7$ fm and $r_{2s}=2.75$ fm 
in the singlet channel \cite{ERE_ref}.
Bound states can be obtained by calculating the position 
of the pole of the $T$ matrix in each channel,
\begin{equation}
k\cot \delta_2(k)=-ik.
\label{pole}
\end{equation}
In the triplet (singlet) channel the positive (negative) scattering length
signals a real (virtual) bound state.
At NLO in free space the deuteron energy is $E^{free}_d=-2.213$ MeV.

When two nucleons are confined within the harmonic trap, 
we diagonalize the Hamiltonian (\ref{hami}) for $A=2$ with the potential
(\ref{V_two_body_eft}) in the basis of HO wavefunctions
$\phi_{n l}(\vec{\xi}_1)$. 
The unperturbed levels are characterized by $N_2=2n+l$.
Since to NLO the inter-nucleon potential is purely
$S$-wave, levels with $l\ge 1$ are unaffected by it.
The $S$-wave energies $E_{2;n}$, on the other hand, depend on
the EFT parameters and thus on the phase shifts.
These energies
are solutions of the transcendental equation 
\begin{equation}
\frac{\Gamma(3/4-E_{2;n}/2 \omega)}{\Gamma(1/4-E_{2;n}/2\omega)}
=-\sqrt{E_{2;n}/2\omega} \; \cot\delta_2\left(\sqrt{m_N E_{2;n}}\right),
\label{eq:scat_2b}
\end{equation}
where $\delta_2(k)$ is given by Eq. (\ref{ERE}).
Equation (\ref{eq:scat_2b}) was first obtained \cite{HT}
by solving the Schr\"odinger 
equation using a pseudopotential \cite{pseudo},
but it can be derived directly within the EFT framework \cite{us1} 
(see also Ref. \cite{mehen}).
As in the absence of the trap, 
at LO the right-hand side contains only the scattering-length term, 
while at NLO the effective-range term appears as well.

\begin{figure}[t]
\begin{center}
%\vspace*{-3cm} 
\includegraphics[scale=0.45,angle=-90,clip=true]{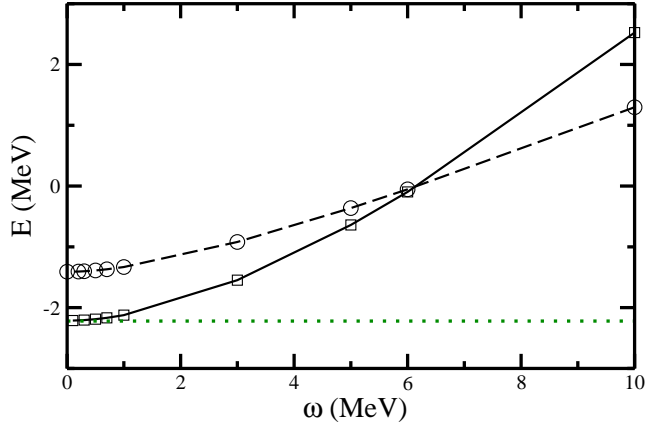}
\end{center} 
\caption{Ground-state energy of the trapped two-nucleon system in the $^3S_1$ 
channel (deuteron in the trap) as a function of the frequency $\omega$. 
The energy at LO (NLO) is given by the dashed (solid) line. 
For small values of $\omega$, the energy converges to the value in free space,
which is, at NLO, 
indicated by the dotted line.}
\label{deuteron_trap}
\end{figure}

As an illustration, we consider the ``deuteron in the trap'', that is,
the lowest state in the trap
which goes into the deuteron as $b\gg a_{2t}$.
Figure \ref{deuteron_trap} shows the lowest energy 
$E_d$ of two nucleons
in the triplet configuration as a function of the frequency $\omega$.
The energy $E_d$ is obtained
by solving Eq. (\ref{eq:scat_2b}). 
At NLO, $E_d=-2.123$ MeV for $\omega=1$ MeV, 
and $E_d=-2.212$ MeV for $\omega=0.1$ MeV,
in good agreement with the energy of the bound state in free space.
Such a good agreement is not surprising, as for small $\omega$ 
the corrections to the energy due to the trap scale in LO with 
$(a_{2t}/b)^4$ \cite{luu}.

%%%%%%%%%%%%%%%%%%%%%%%%%%%%%%%%%%%%%%%%%%%%%%%%%%%%%%%%%% 
%%%%%%%%%%%%%%%%%%%%%%%%%%%%%%%%%%%%%%%%%%%%%%%%%%%%%%%%%% 

\begin{figure}[tb]
\begin{center}
%\vspace*{-3cm} 
\includegraphics[scale=0.33,angle=-90,clip=true]{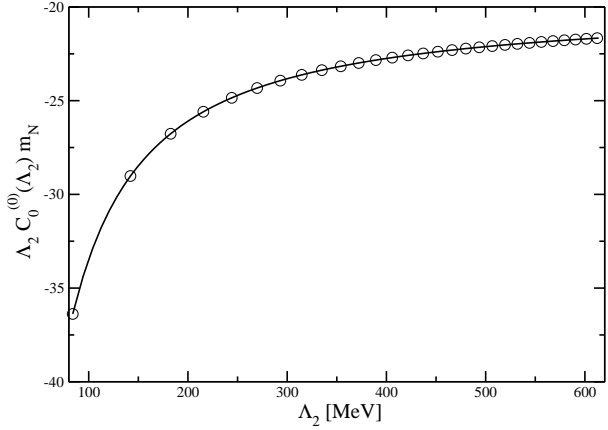}
\end{center} 
\caption{LO two-nucleon coupling constant 
$\Lambda_2 C_{0}^{(0)}(\Lambda_2)m_N$ in 
the $^3S_1$ channel as a function 
of the cutoff $\Lambda_2$, for $\omega=1$ MeV.}
\label{cc_LO}
\end{figure}

\begin{figure}[t]
\begin{minipage}[t]{75mm}
\centerline{\includegraphics[scale=0.33,angle=-90,clip=true]{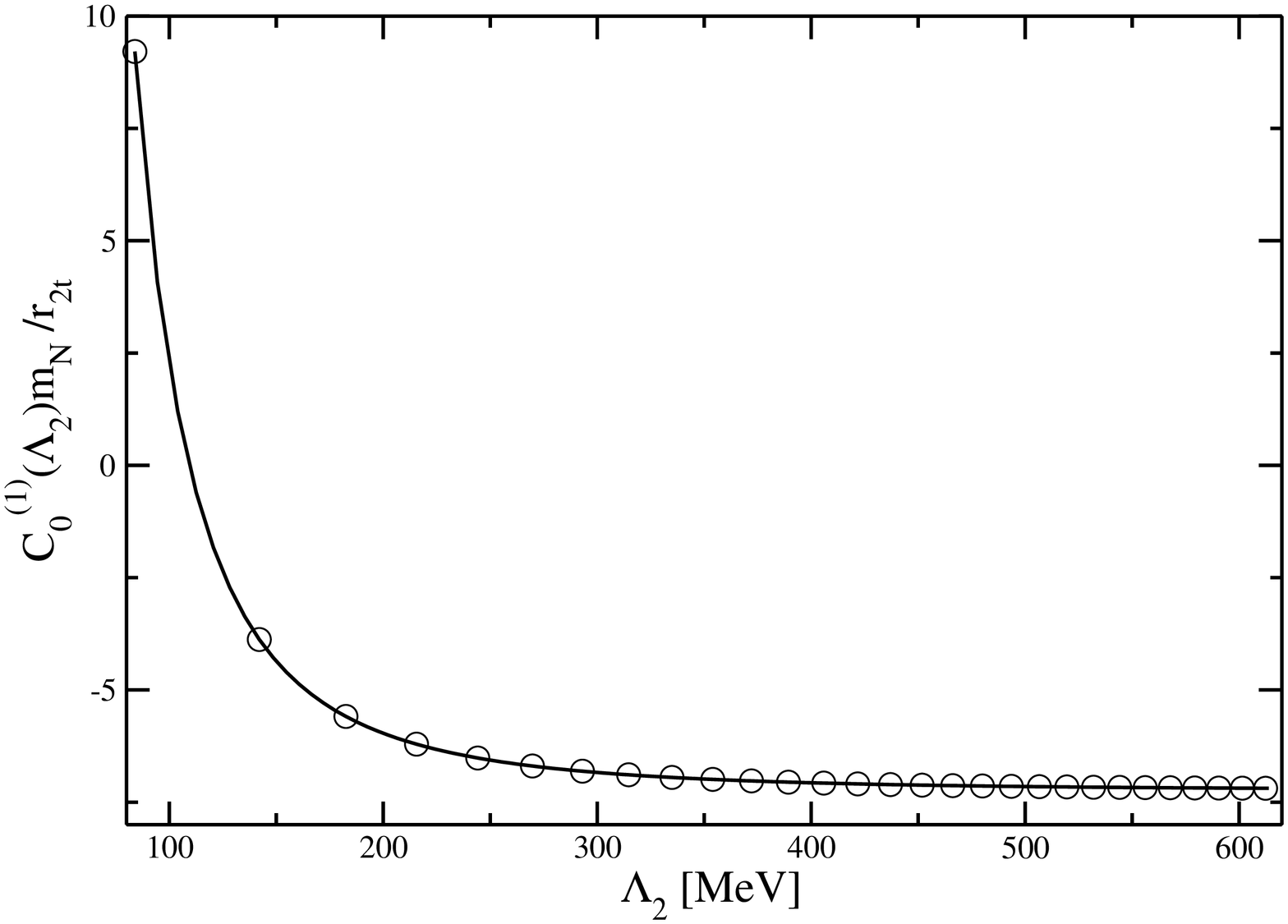}}
\end{minipage}
\hspace{\fill}
\begin{minipage}[t]{80mm}
\centerline{\includegraphics[scale=0.33,angle=-90,clip=true]{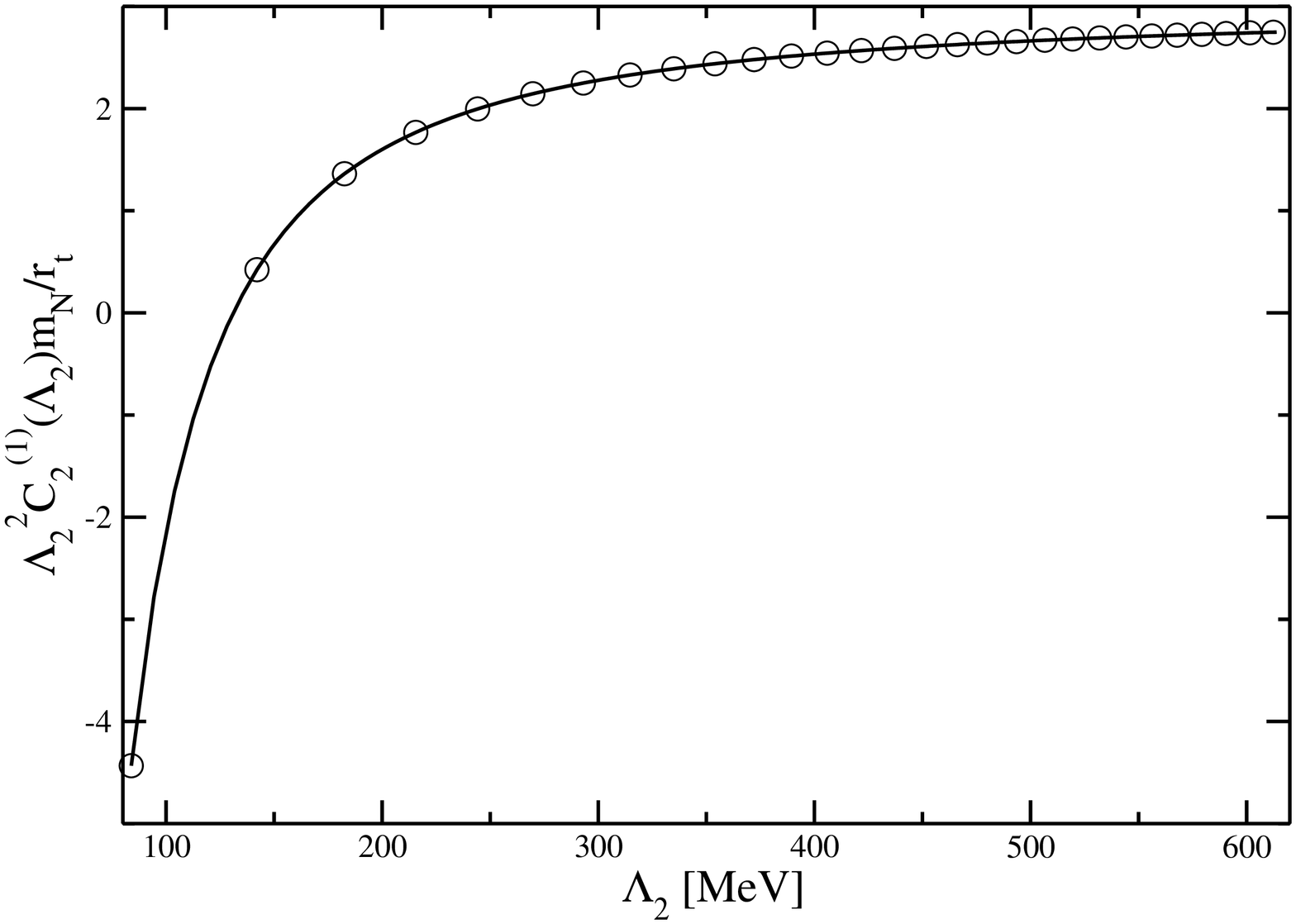}}
\end{minipage}
\caption{NLO two-nucleon coupling constants 
in the $^3S_1$ channel as a function 
of the cutoff $\Lambda_2$, for $\omega=1$ MeV:
$C_0^{(1)}(\Lambda_2) m_N/r_{2t}$ (left panel) and
$\Lambda_2^2 C_{2}^{(1)}m_N/r_{2t}$ (right panel).}
\label{cc_NLO}
\end{figure}

We use this and other energy levels given by Eq. (\ref{eq:scat_2b}) 
to determine the parameters $C_0$, $C_2$, {\it etc.}
In the triplet channel, $C_{0}^{(0)}(\Lambda_2)$ is found by demanding that
it produces the LO deuteron in the trap.
In Fig. \ref{cc_LO} we show the running of the triplet coupling
constant $\Lambda_2 C_{0}^{(0)}(\Lambda_2)m_N$  
as a function of the cutoff $\Lambda_2$, for $\omega=1$ MeV.
For large $\Lambda_2$, 
$C_{0}^{(0)}(\Lambda_2)\rightarrow -2 \pi^2 (m_N \Lambda_2)^{-1}$.
At NLO, $C_{0}^{(1)}(\Lambda_2)$ and $C_{2}^{(1)}(\Lambda_2)$
are obtained from the NLO deuteron in the trap and the first
excited-state solution of Eq. (\ref{eq:scat_2b}) 
considering both the scattering length $a_{2t}$ 
and the effective range $r_{2t}$.
In Fig. \ref{cc_NLO} 
we show the running of the triplet coupling
constants 
$C_0^{(1)}(\Lambda_2) m_N/r_{2t}$
and $\Lambda_2^2 C_{2}^{(1)}(\Lambda_2)m_N/r_{2t}$
as a function of the cutoff  $\Lambda_2$, again for $\omega=1$ MeV.
For large enough $\Lambda_2$, 
$C_0^{(1)}(\Lambda_2)$ becomes constant 
and $C_{2}^{(1)}(\Lambda_2) \propto \Lambda_2^{-2}$ \cite {us1}.
The running of coupling constants is qualitatively
similar for other frequencies.
In the singlet channel we obtain analogous results using as input 
the corresponding scattering parameters. 

%%%%%%%%%%%%%%%%%%%%%%%%%%%%%%%%%%%%%%%%%%%%%%%%%%%%%%%%%%%%%%
%%%%%%%%%%%%%%%%%%%%%%%%%%%%%%%%%%%%%%%%%%%%%%%%%%%%%%%%%%%%%%

%%%%%%%%%%%%%%%%%%%%%%%%%%%%%%%%%%%%%%%%%%%%%%%%%%%%%%%%%%%%%%
%%%%%%%%%%%%%%%%%%%%%%%%%%%%%%%%%%%%%%%%%%%%%%%%%%%%%%%%%%%%%%

\begin{figure}[t]
\begin{center}
%\vspace*{-3cm} 
\includegraphics[scale=0.33,angle=-90,clip=true]{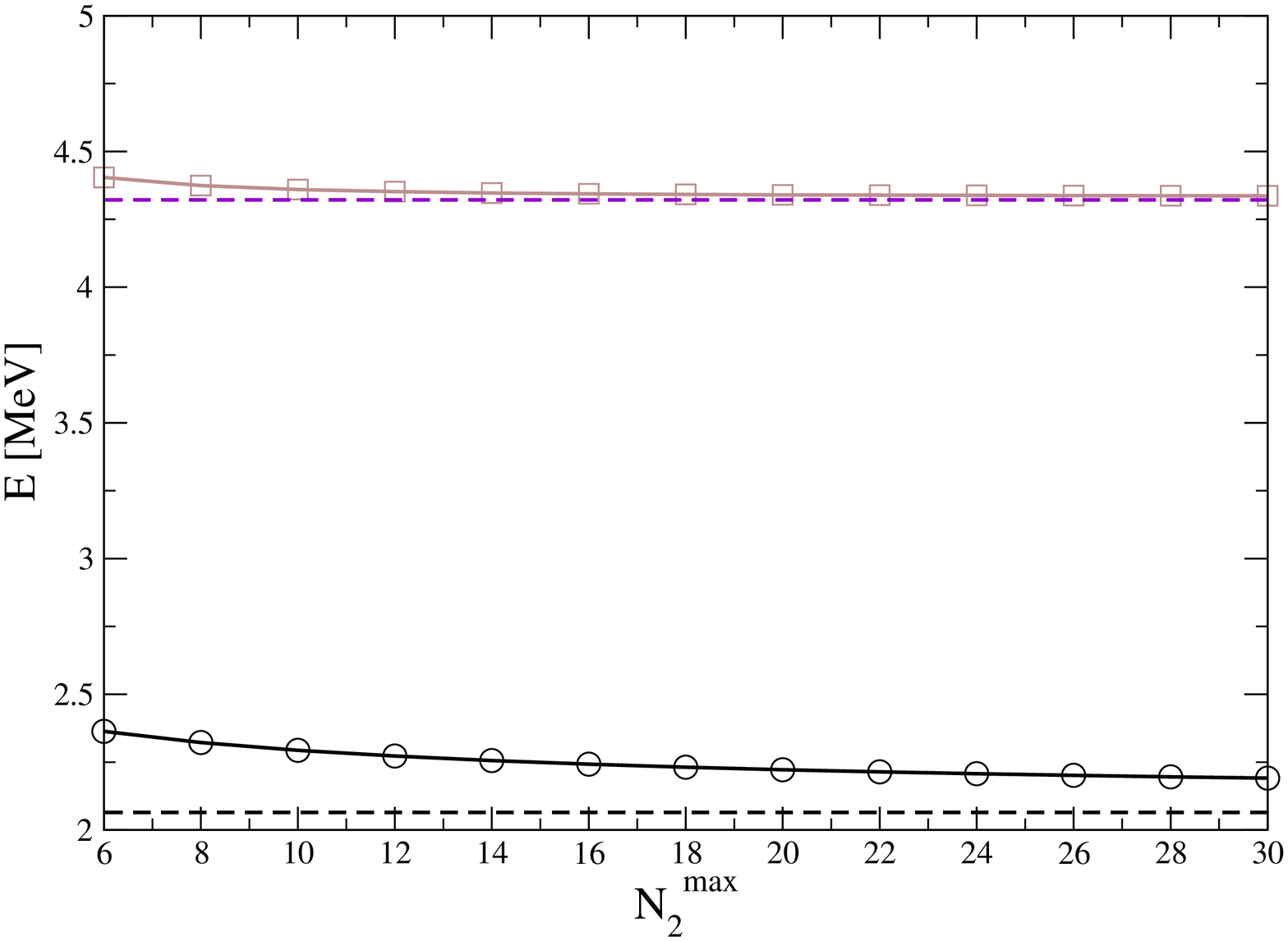}
\end{center} 
\caption{Energies of the first (circle) and second (square) excited states 
of the two-nucleon system in the $^3S_1$ channel at LO and NLO, respectively, 
as function of $N_2^{max}$,
for $\omega=1$ MeV.
The dashed lines correspond to the exact solution given by 
Eq. (\ref{eq:scat_2b}).}
\label{deuteron_exc}
\end{figure}

Once the EFT couplings are determined from a couple of levels,
the other levels can be calculated. They do not agree exactly
with the levels of Eq. (\ref{eq:scat_2b}) at finite $\Lambda_2$,
but approach them as $\Lambda_2\to \infty$.
In Fig. \ref{deuteron_exc}  we show the first excited state at LO and the 
second excited state at NLO as a function of the size of the model space 
characterized by $N_2^{max}$. 
As $N_2^{max}$
increases the energies in the finite model space converge to the exact 
energies of Eq. (\ref{eq:scat_2b}).
It is clear that convergence is sped up when the correction 
to the potential at NLO is included: the first excited state is 
now simply fitted,
and the second excited state is very close to the exact value
even at relatively small 
$N_2^{max}$.

%%%%%%%%%%%%%%%%%%%%%%%%%%%%%%%%%%%%%%%%%%%%%%%%%%%%%%%%%%%%%%
%%%%%%%%%%%%%%%%%%%%%%%%%%%%%%%%%%%%%%%%%%%%%%%%%%%%%%%%%%%%%%

\begin{figure}[tb]
\begin{minipage}[t]{75mm}
\centerline{\includegraphics[scale=0.33,angle=-90,clip=true]{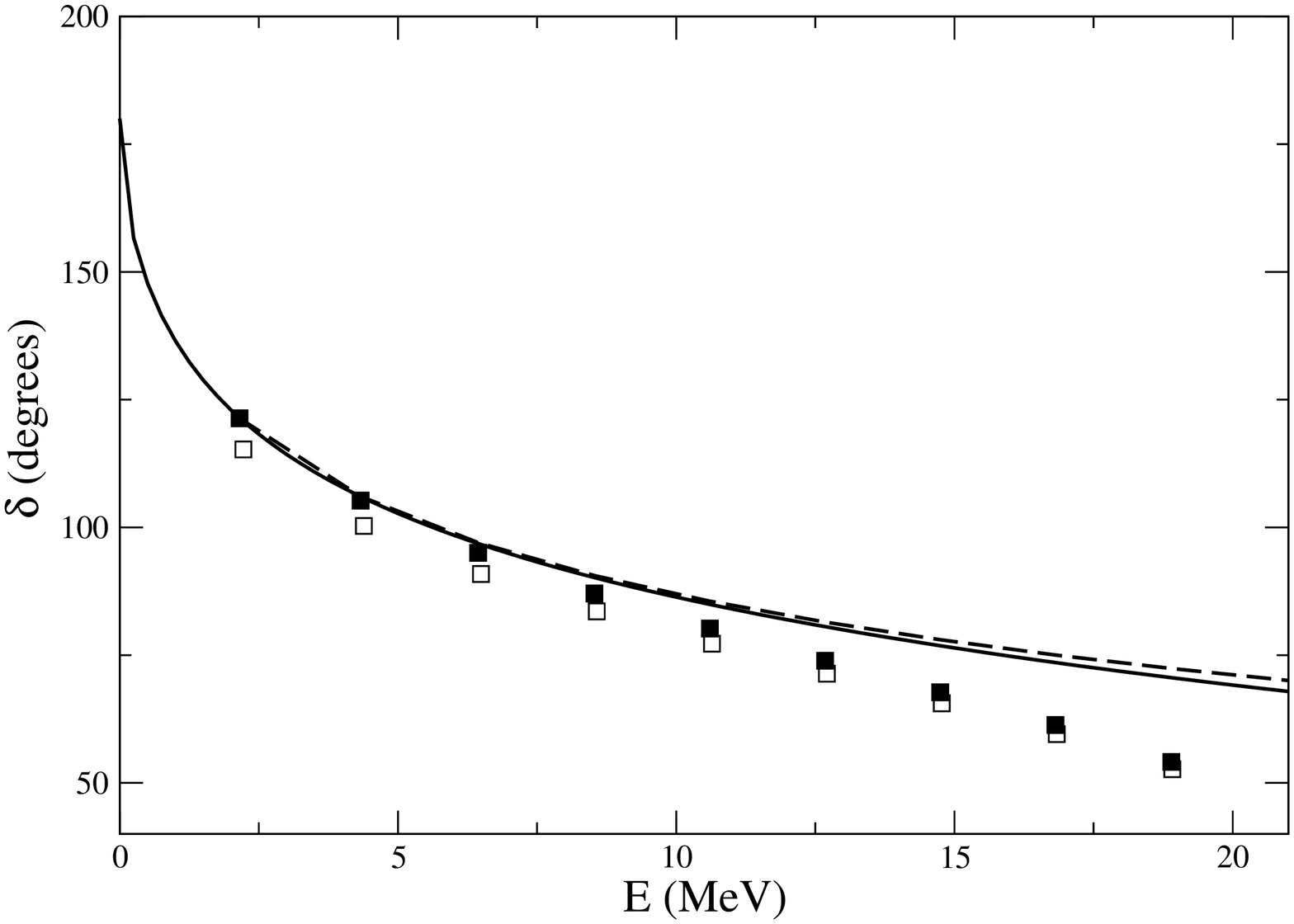}}
\end{minipage}
\hspace{\fill}
\begin{minipage}[t]{80mm}
\centerline{\includegraphics[scale=0.33,angle=-90,clip=true]{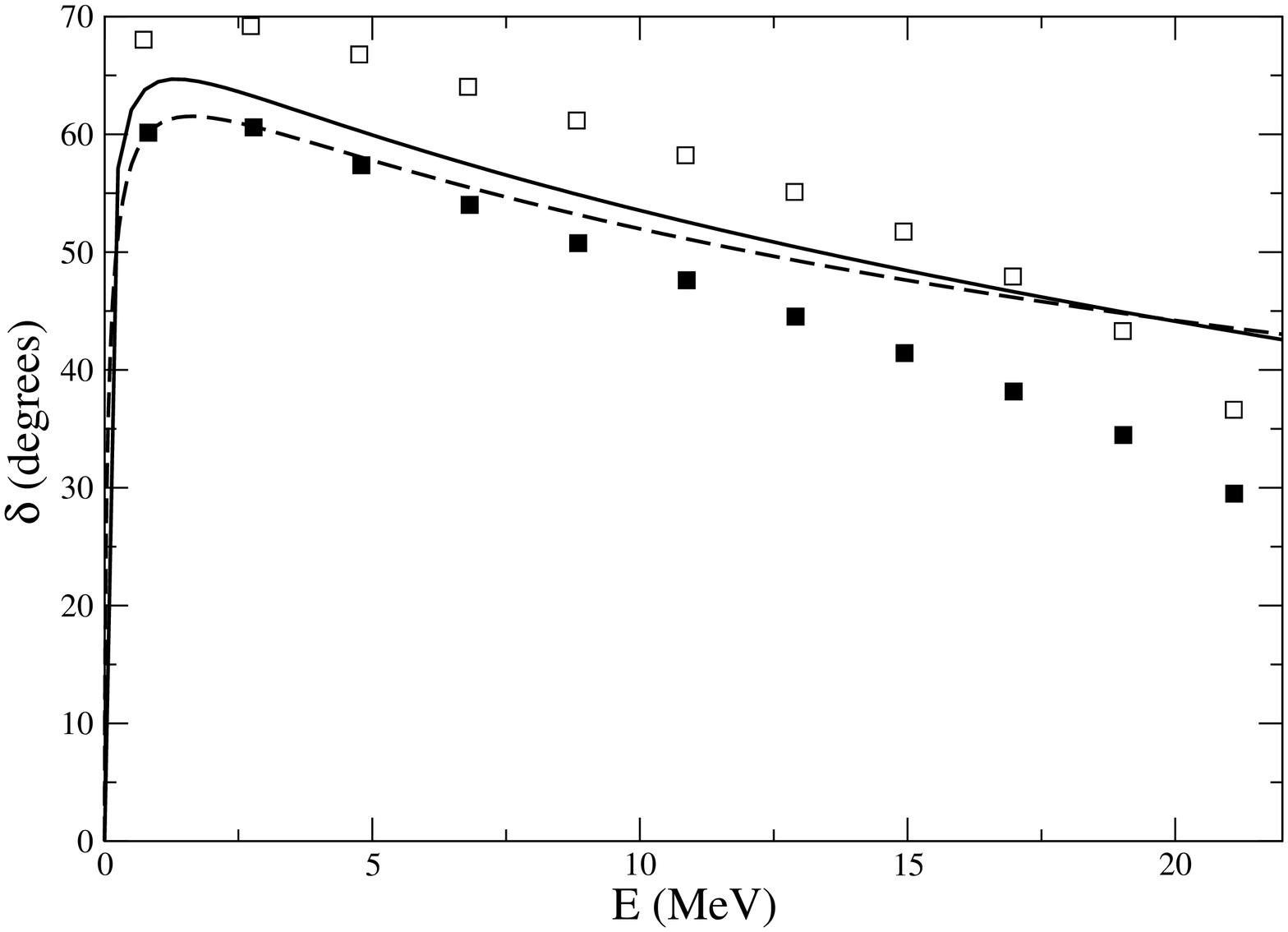}}
\end{minipage}
\caption{
%Singlet and triplet 
Phase shifts for the two-nucleon 
system at $\omega=1$ MeV and $N_{max}=20$ as a function of the relative 
energy:
$^3S_1$ (left panel) and $^1S_0$ (right panel). 
EFT results at LO (NLO) are marked by empty (filled) squares;
the ERE up to the effective range is indicated by a dashed line; 
and the Nijmegen $np$ PSA \cite{nijmegen} by a solid line. 
}
\label{fig:phaseshiftNLO}
\end{figure}

As discussed in Ref. \cite{us1}, 
with the calculated levels input 
to the left-hand side of Eq. (\ref{eq:scat_2b})
we can invert the procedure and obtain
scattering phase shifts for a given cutoff. 
In Fig. \ref{fig:phaseshiftNLO}, we plot the phase shifts for both 
triplet and singlet configurations,
for $\omega=1$ MeV and $N_{max}=20$.
Also displayed are
the corresponding ERE phase shifts 
and
the Nijmegen neutron-proton ($np$)
phase-shift analysis (PSA) \cite{nijmegen}.
(The discrepancy observed in the $^1S_0$ channel between ERE
and PSA phase shifts appears because the ERE is calculated with 
neutron-neutron scattering length and effective range.
The difference is an isospin-breaking effect of higher order than 
what is considered here.)
At low energies, as expected, 
one obtains good agreement with the ERE and with 
the Nijmegen PSA. 
Agreement worsens as one goes to higher and higher energies, 
since the higher energy levels show more effect of the finite cutoff,
which effectively induces higher-order ERE terms.
Better agreement with ERE is obtained as the cutoff increases,
and, for a given cutoff, agreement
improves systematically order by order, as long as the momentum of the state 
is well below the cutoff imposed by the model space 
(see Fig. 5 of Ref. \cite{us1}).

%%%%%%%%%%%%%%%%%%%%%%%%%%%%%%%%%%%%%%%%%%%%%%%%%%%%%%%%%% 
%%%%%%%%%%%%%%%%%%%%%%%%%%%%%%%%%%%%%%%%%%%%%%%%%%%%%%%%%%%%%%

One can look at the phase shifts as predictions of the theory,
but there are easier ways to carry out 
this two-nucleon calculation (see for example Ref. \cite{aleph}).
The motivation to use the HO basis comes from wanting to study larger systems.
We now turn to the simplest such cases, the three-nucleon system.

\section {Three nucleons in a harmonic trap }
\label{sec_3n}

We now consider a system of three nucleons trapped within an 
HO potential. The situation here involves elements
encountered before in both the two-state-fermion \cite{trap0,us2}
and boson \cite{hanstrap} cases.

Since no symmetry forbids a three-nucleon potential, we have
to consider this additional element.
In pionless EFT, the three-nucleon potential
is also expressed in terms of delta functions, 
\begin{equation}
V(\vec{\xi_1}, \vec{\xi_2})= 
D_0 \, \delta(\vec{\xi_1}) \, \delta(\vec{\xi_2}) +\ldots,
\label{3BF}
\end{equation}
where $D_0$ is a parameter and derivatives terms are buried in ``\ldots''.
Just as the two-body parameters, $D_0$ 
and other three-body parameters depend on the cutoff.
In Ref. \cite{pionlessT} it has been
shown that in free space 
the non-derivative three-body force 
is needed for RG invariance already at LO,
while derivative corrections appear at NNLO and higher.
The non-derivative $D_0$ interaction 
contributes only when we place three nucleons at the origin
in a relative $S$ wave,
that is, when the total isospin is $T=1/2$ and the 
total angular momentum and parity is $J^{\pi}=1/2^{+}$.
To NLO, we thus need information about one three-nucleon observable,
such as the ground-state energy.

In the trap, we use
the basis states
\begin{equation}
{\cal A}\left\{\left[\phi_{n_1 l_1}(\vec{\xi}_1) \otimes 
\phi_{n_2 l_2}(\vec{\xi}_2) \right]_{L} 
|(\frac{1}{2} \frac{1}{2} )s_2 \frac{1}{2};S\rangle
|(\frac{1}{2} \frac{1}{2} )t_2 \frac{1}{2};T\rangle
\right\},
\label{eq:basis3b}
\end{equation} 
which have the spatial part constructed using HO wavefunctions in 
$\vec{\xi}_1$ and $\vec{\xi}_2$
with quantum numbers  $n_1$, $l_1$ and 
$n_2$, $l_2$, respectively, 
with the angular momentum coupled to $L$, while the spin (isospin) part 
is constructed by coupling 
first two 
spins (isospins) $s=1/2$ ($t=1/2$) into spin (isospin) $s_2$  ($t_2$) 
and then a third spin (isospin) $s=1/2$ ($t=1/2$)
to total spin $S$ (isospin $T$).
In Eq.~(\ref{eq:basis3b}), ${\cal A}$ stands for the operator that 
antisymmetrizes the three-particle wavefunction.  
Details on the construction of a fully antisymmetrized basis can be found 
in Ref. \cite{petr}. 
The basis states thus constructed are eigenstates of the 
unperturbed Hamiltonian $H_0$ 
with $N_3=2(n_1+n_2)+l_1+l_2$.

In the conventional NCSM approach, it is customary to choose the truncation 
in the two-body system so that the many-body space is the minimal required to 
include completely the two-body space. For example, if we consider just 
$S$-wave interactions, $N^{max}_2=N^{max}_3$ when one describes positive-parity
states.
However, one has to consider that the renormalization of the two-body system
means that states lying above the cutoff $N_2^{max}$
have been ``integrated out" rather than simply discarded. Their effects 
are, thus, included implicitly in the effective two-body interaction. 
When these two interacting particles are embedded in a system with a 
larger number of particles, the spectators will carry energies associated 
with the HO levels they occupy. For example, of the $(N_3+3)\omega$ total 
energy of one of the basis states (\ref{eq:basis3b}), 
$(2n_2+l_2+3/2)\omega$ 
is carried by the relative motion of the 
spectator. As a consequence, the maximum energy available to the two-body
subsystem is smaller than that allowed by the three-body cutoff $N^{max}_3$
and some of the states removed by the truncation will not be accounted
for by the renormalization. In order to account for all the two-body physics 
beyond our cutoff, we simply decouple the 
cutoff of the many-body problem from that of the subcluster 
defining any interaction.
Each of our calculations is characterized by two cutoff parameters: 
$N^{max}_2$ for the two-body subsystem, 
and $N^{max}_3$ for the three-body system. 
For fixed $N^{max}_2$ and $N^{max}_3$, we calculate the 
three-body energies $E_{3;n}$.
We first increase $N^{max}_3$ till convergence,
which to a good approximation happens already when
$N^{max}_3$ is a couple of units larger than $N^{max}_2$,
and we then increase $N^{max}_2$.
We have shown in a previous publication \cite{us2} that proceeding this way
greatly improves the convergence of the energies 
of a two-state-fermion system.

In the rest of this paper,
we illustrate the application of 
our formalism to nucleons in the two channels with $T=1/2$: 
$J^{\pi}=1/2^{+}, 3/2^{+}$.
These are the most interesting channels,
since they are accessible in nucleon-deuteron scattering.

The energy
in the trap should approach the energy of the untrapped system when
the HO trap is weak, {\it{i.e.}}, for small values of $\omega$. 
The ground state in the $T=1/2$, $J^{\pi}=1/2^{+}$ channel
becomes the triton ($^3$H) 
at an energy $E_{t}=-8.482$ MeV \cite{tritonBE}, 
while all other states coalesce into the continuum
states.
Some of these states 
correspond to the $S$-wave scattering
of a neutron on the deuteron, which can form inside the trap
in the $T=1/2$ channels.
We can then extract the $nd$ phase shifts 
from the three-nucleon energies above the deuteron ground state,
$E_{3;n}-E_{d}$, 
with a relation similar to Eq. (\ref{eq:scat_2b}),
\begin{equation}
 \frac{\Gamma(3/4-(E_{3;n}-E_{d})/2\omega)}
      {\Gamma(1/4-(E_{3;n}-E_{d})/2\omega)}
=-\sqrt{\left(E_{3;n}-E_d\right)/2\omega} 
\; \cot\delta_3\left(\sqrt{2\mu_{nd} \left(E_{3;n}-E_d\right)}\right),
\label{eq:scat_3b}
\end{equation}
where $\mu_{nd}$ is the neutron-deuteron reduced mass,
and the phase shift $\delta_3(k)$ is given by an ERE expansion,
\begin{equation}
k \, \cot \delta_3(k) =-\frac{1}{a_3}+\frac{r_3}{2}k^{2}+ \ldots
\label{ERE3}
\end{equation}
in terms of $nd$ ERE parameters $a_3$, $r_3$, {\it etc.}
In the $S=3/2$ and $S=1/2$ channels
the experimental values \cite{d_n_exp} of 
the scattering lengths are 
$a_{3q}=6.35 \pm 0.02 ~\rm{fm}$ 
and $a_{3d}=0.65\pm0.04$ fm,
respectively.

It is important to note that Eq. (\ref{eq:scat_3b}) holds as long as 
the range of the $nd$ interaction is much smaller than the 
effective trapping length 
$b'=1/\sqrt{\mu_{nd}\omega}$.
This makes high-cutoff calculations 
challenging since the $nd$ size 
is rather large, of the order of the triplet two-nucleon scattering length. 
In the two-state fermion case we did succeed in extracting
the atom-dimer scattering length using this method \cite{us2}.

\subsection{The channel $T=1/2$, $J^{\pi}=3/2^{+}$}
\label{sec_3_half}

We first consider the case of three nucleons with $T=1/2$ 
coupled to $J^{\pi}=3/2^{+}$. 
In this channel the three-nucleon force appears only in higher orders
than included here, so the properties of the three-nucleon
system are determined by the two-nucleon input.
This situation is the same as for three 
two-state fermions \cite{trap0,us2}.

\begin{figure}[tb]
\begin{minipage}[t]{75mm}
\centerline{\includegraphics[scale=0.33,angle=-90,clip=true]{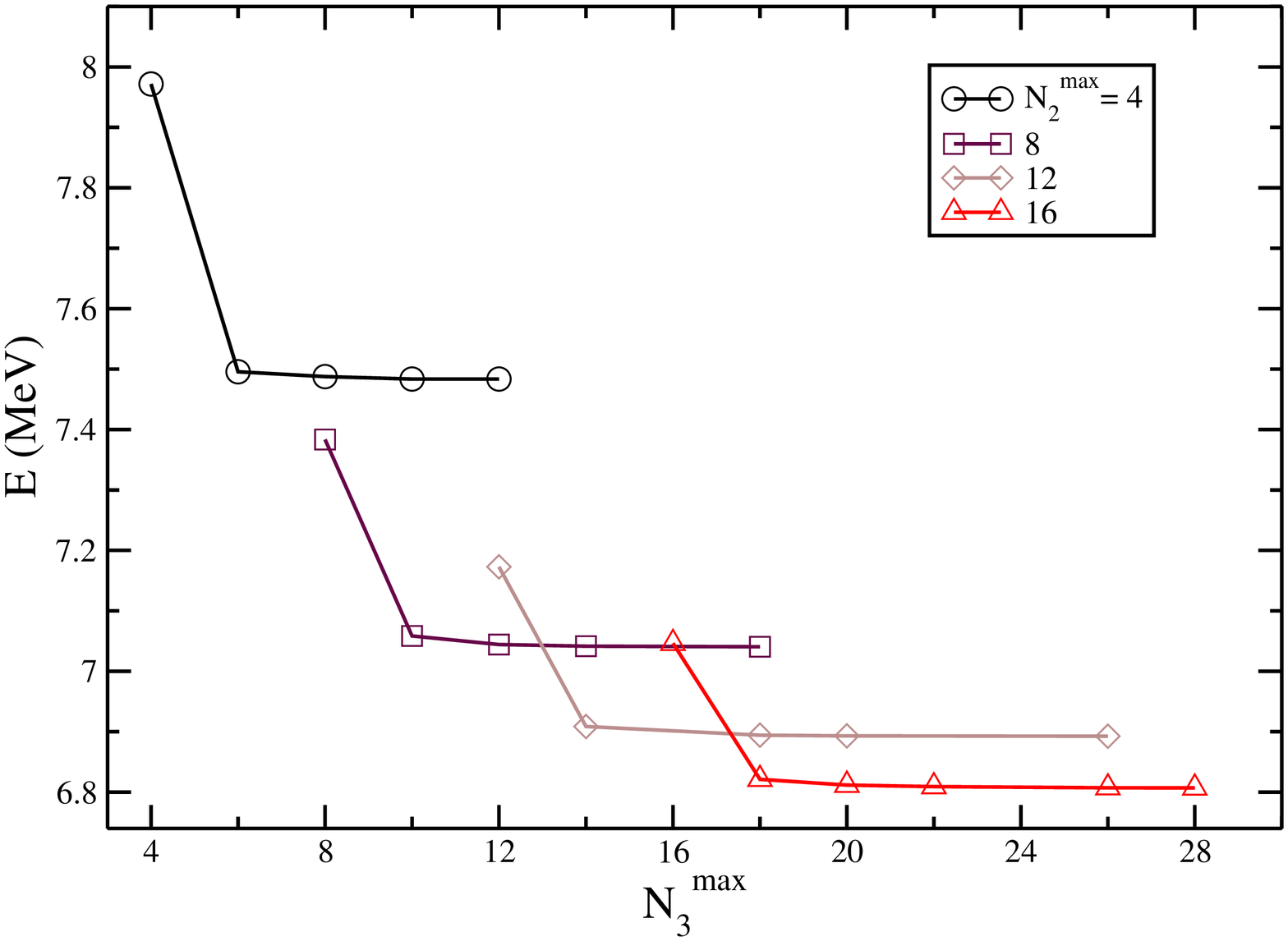}}
\end{minipage}
\hspace{\fill}
\begin{minipage}[t]{80mm}
\centerline{\includegraphics[scale=0.33,angle=-90,clip=true]{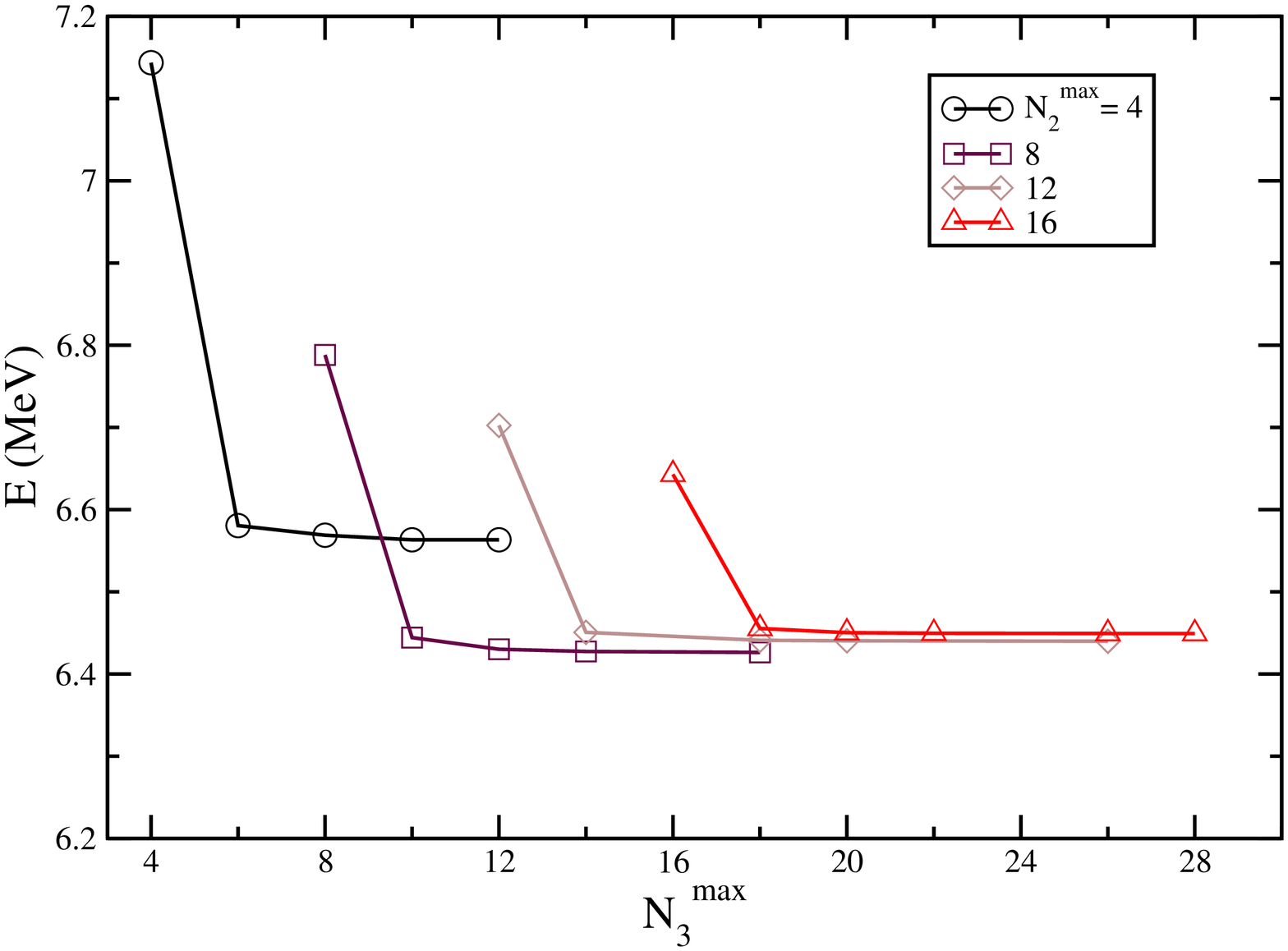}}
\end{minipage}
\caption{
Ground-state energy of the trapped three-nucleon system 
coupled to $T=1/2$, $J^{\pi}=3/2^{+}$
as function of the three-body model-space size $N_{3}^{max}$,
for $\omega=3$ MeV: LO (left panel) and NLO (right panel). 
Results 
are shown for different values of the two-body model-space size $N_2^{max}$.
}
\label{three_3_half_LO&NLO}
\end{figure}

We start by discussing the convergence of the energy levels.
For illustration, we take the ground state at a relatively
large frequency $\omega =3$ MeV, but qualitative features
are the same for other states and frequencies.
We show LO and NLO results in Fig. \ref{three_3_half_LO&NLO}
for various values of the two-body model space size $N_2^{max}$. 
For each $N_2^{max}$, the three-body model-space size 
defined by $N_3^{max}$ is increased.
There is a sharp decrease of the energy as $N_3^{max}=N_2^{max}+2$
and, as $N_3^{max}$ increases further,
the ground-state energy reaches a converged value. 
More precisely, as $N_3^{max}$  increases, the energy 
changes by less than 0.1 keV for the values considered here. 
Thus faster convergence is obtained for a given $N_2^{max}$ by increasing
$N_3^{max}$ beyond $N_2^{max}$, as observed before 
in the case of two-state fermions \cite{us2}. 
This can be seen for instance by comparing the LO energy obtained 
for $N_{2}^{max}=8$ and $N_3^{max}=10$, which is
$E_3^{LO}=7.058$ MeV, and the LO energy for 
$N_{2}^{max}=N_3^{max}=12$, which is $E_3^{LO}=7.173$ MeV
and further away from the extrapolated value
$E_3^{LO}(\infty)$ obtained from  Eq. (\ref{eq:fitE}) below.

%%%%%%%%%%%%%%%%%%%%%%%%%%%%%%%%%%%%%%%%%%%%%%%%%%%%%%%%%% 
%%%%%%%%%%%%%%%%%%%%%%%%%%%%%%%%%%%%%%%%%%%%%%%%%%%%%%%%%%%%%%

\begin{figure}[t]
\begin{center}
%\vspace*{-3cm} 
\includegraphics[scale=0.33,angle=-90,clip=true]{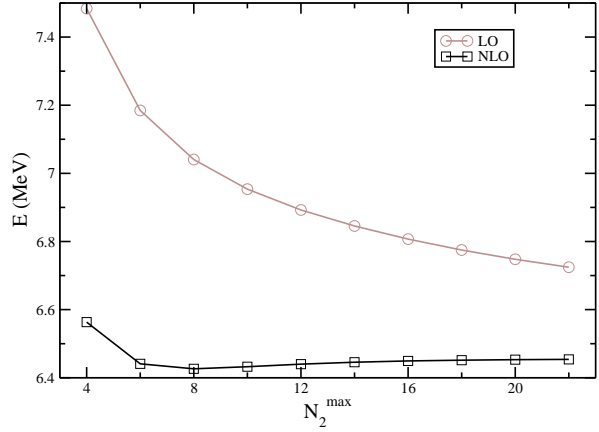}
\end{center} 
\caption{Ground-state energy of the trapped
three-nucleon system coupled to $T=1/2$ , $J^{\pi}=3/2^+$
as a function of $N_2^{max}$,
for $\omega= 3$ MeV: LO (circles) and NLO (squares).}
\label{three_3_half_LO_NLO}
\end{figure}

Figure \ref{three_3_half_LO_NLO} shows the convergence with respect to 
$N_2^{max}$. 
Clearly the energy converges to a finite value as the two-body cutoff 
increases. We can thus confirm that, as in the free-space case 
\cite{d_n_eft}, no three-nucleon force is needed at these orders 
to renormalize the three-nucleon system.
We can fit the cutoff dependence of the energy  
with 
\begin{equation}
E_3(N_2^{max})=E_3(\infty)
+\frac{\epsilon_1}{(N_2^{max}+3/2)^{1/2}}
+\frac{\epsilon_3}{(N_2^{max}+3/2)^{3/2}},
\label{eq:fitE}
\end{equation}
where $E_3(\infty)$ is the asymptotic value and 
$\epsilon_{1,3}$ give the rate of convergence. 
The fits are performed for $N_2^{max} \geq 12$.
At LO we obtain $\epsilon_1^{LO}= 2.270$ MeV, $\epsilon_3^{LO}=1.676$ MeV,
and $E^{LO}_3(\infty)=6.241$ MeV, which is $\sim 500$ keV 
below the value obtained with the largest considered cutoff, $N_2^{max}=22$.
At NLO, one obtains instead
$\epsilon_1^{NLO}= 0.3$ MeV, $\epsilon_3^{NLO}=-2.99$ MeV, 
and $E^{NLO}_3(\infty)=6.417$ MeV, which is only $\sim 40$ keV 
above the value obtained at $N_2^{max}=22$. The convergence at NLO is faster 
than that at LO, as can be seen in Fig. \ref{three_3_half_LO_NLO}
and in the decrease of the coefficient $\epsilon_1$ of the leading cutoff
error.
Although the specific numbers above depend strongly on
$\omega$, the convergence pattern is qualitatively the same
for other values of $\omega$.

%%%%%%%%%%%%%%%%%%%%%%%%%%%%%%%%%%%%%%%%%%%%%%%%%%%%%%%%%%%%%%
%%%%%%%%%%%%%%%%%%%%%%%%%%%%%%%%%%%%%%%%%%%%%%%%%%%%%%%%%%%%%%

\begin{figure}[tb]
\begin{minipage}[t]{75mm}
\centerline{\includegraphics[scale=0.33,angle=-90,clip=true]{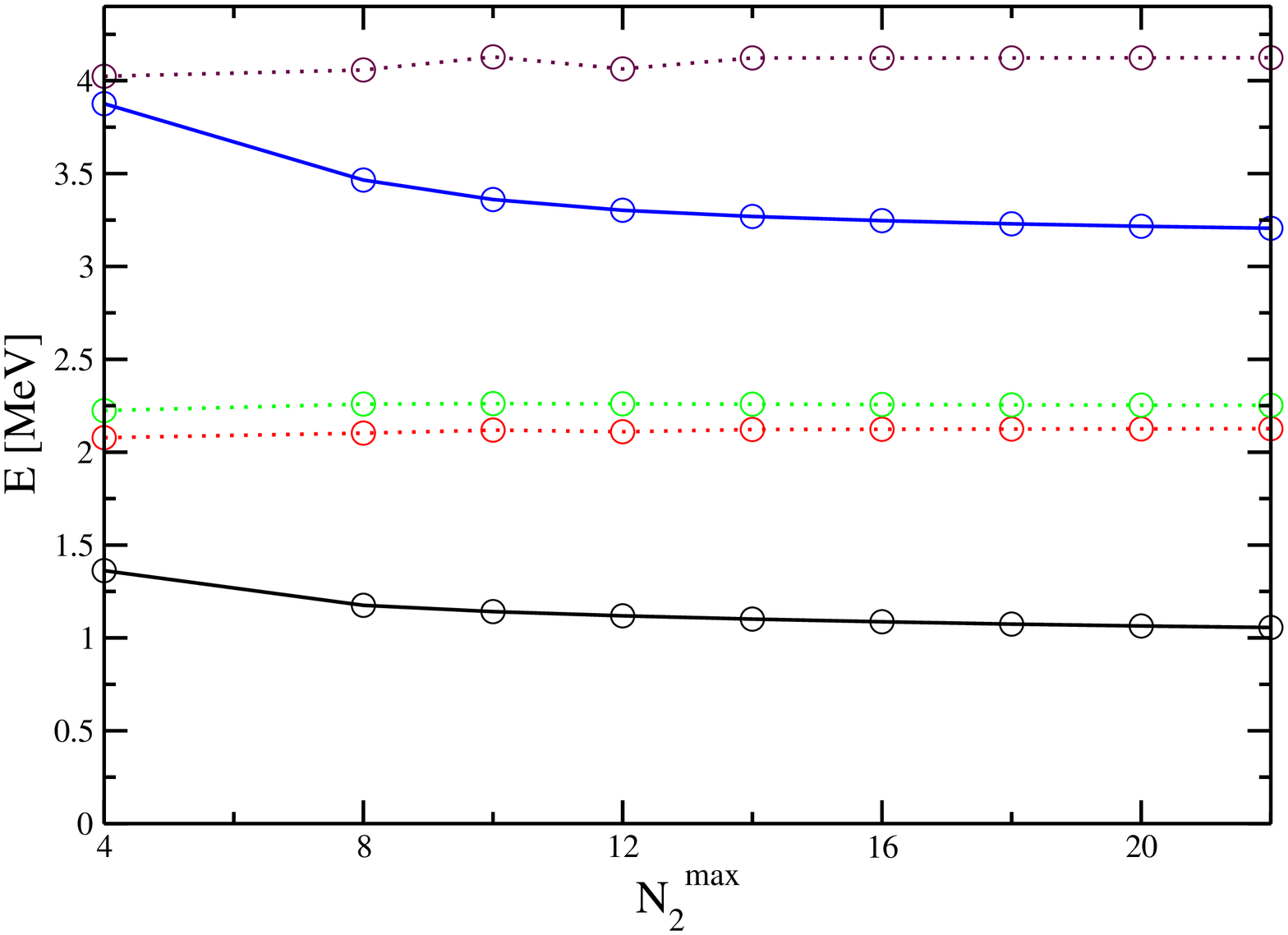}}
\end{minipage}
\hspace{\fill}
\begin{minipage}[t]{80mm}
\centerline{\includegraphics[scale=0.33,angle=-90,clip=true]{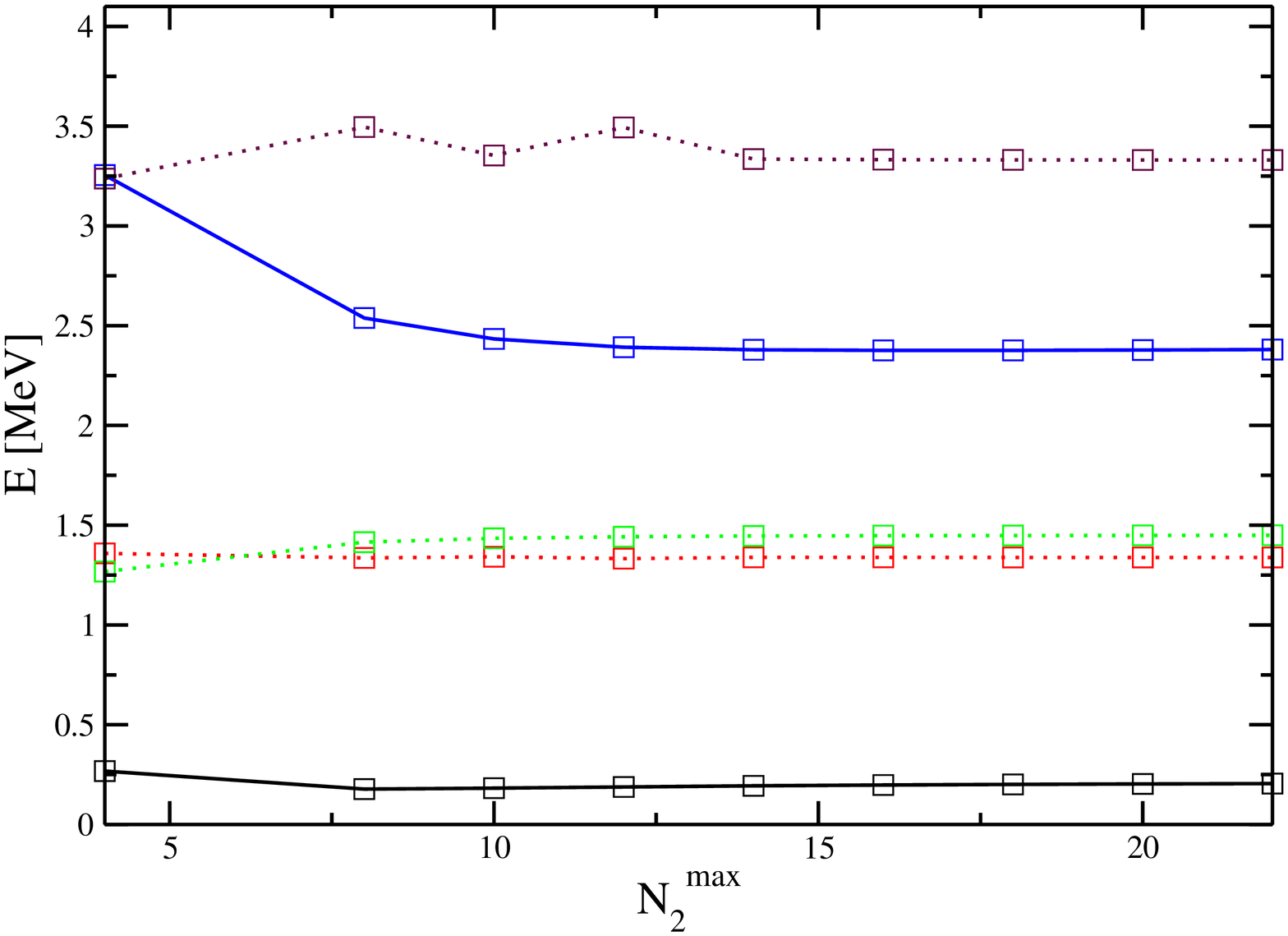}}
\end{minipage}
\caption{
Lowest energies of the trapped three-nucleon system
with $T=1/2$, $J^{\pi}=3/2^{+}$ as a function of $N_2^{max}$, for 
$\omega=1$ MeV: LO (left panel) and NLO (right panel).
The ground state and the third excited state (full lines) correspond to 
neutron-deuteron scattering within the trap
in the $L=0$, $S=3/2$ channel,
whereas the other states shown correspond to different $L$, $S$ 
configurations. 
}
\label{states_LO&NLO_d_n}
\end{figure}

For $\omega=1$ MeV, the first 
few eigenstates characterized by $J^{\pi}=3/2^{+}$ and  $T=1/2$ 
are shown in Fig. \ref{states_LO&NLO_d_n}.
Since there is no free-space three-nucleon bound state in this channel,
the lowest eigenstates of Eq. (\ref{hami}) for a weak trap correspond to 
``discretized'' $nd$ scattering states confined within the trap. 
For $nd$ scattering in a $S$ wave, 
we can select the lowest eigenstates with the configuration 
of orbital angular momentum $L=0$ and spin
$S=3/2$,
where the two-nucleon interaction in the $^1S_0$ channel does 
not play any role.

%%%%%%%%%%%%%%%%%%%%%%%%%%%%%%%%%%%%%%%%%%%%%%%%%%%%%%%%%% 
%%%%%%%%%%%%%%%%%%%%%%%%%%%%%%%%%%%%%%%%%%%%%%%%%%%%%%%%%%%%%%

\begin{figure}[t]
\begin{center}
\includegraphics[scale=0.33,angle=-90,clip=true]{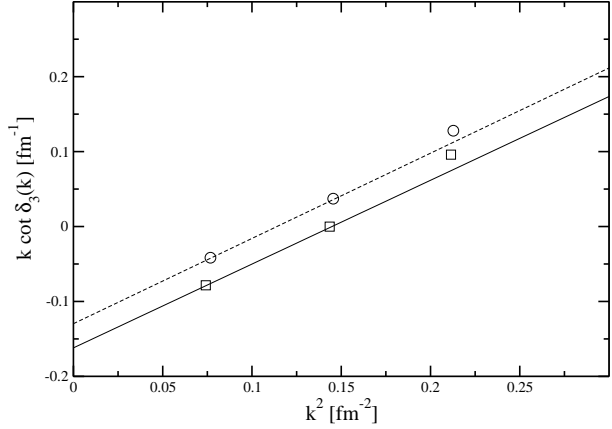}
\end{center} 
\caption{ $k \cot \delta_3(k)$ in the $T=1/2$, $L=0$, $S=3/2$ channel obtained 
from the lowest energies at LO (circles) and NLO (squares)
as a function of $k^2$,
for $\omega=1$ MeV and $N_2^{max}=18$. 
The dashed (full) line corresponds to a LO (NLO) linear fit
to the two lowest energies.
} 
\label{quartetphases}
\end{figure}

{}From the lowest energies of the eigenstates of the  
Hamiltonian ($\ref{hami}$) in the $L=0$, $S=3/2$ configuration we extract 
the $S$-wave $nd$ phase shifts using Eq. (\ref{eq:scat_3b}). 
In Fig. \ref{quartetphases} we show $k \cot\delta_3(k)$ for 
$\omega=1$ MeV and $N_2^{max}=18$.
We extract the value of the quartet scattering length $a_{3q}$ 
using Eq. (\ref{ERE3}).  
{}From the two lowest energies we obtain: $a_{3q}^{LO}=7.71$ fm 
and $a_{3q}^{NLO}=6.30$ fm.
Note that by considering the second and third phase-shift points 
in Fig. \ref{quartetphases}
we would get instead: $a_{3q}^{LO}=6.29$ fm and $a_{3q}^{NLO}=4.92$ fm,
which shows that
higher-order ERE terms are more important at higher energy.

%%%%%%%%%%%%%%%%%%%%%%%%%%%%%%%%%%%%%%%%%%%%%%%%%%%%%%%%%%%%%%
%%%%%%%%%%%%%%%%%%%%%%%%%%%%%%%%%%%%%%%%%%%%%%%%%%%%%%%%%%%%%%

\begin{figure}[t]
\begin{center}
\includegraphics[scale=0.33,angle=-90,clip=true]{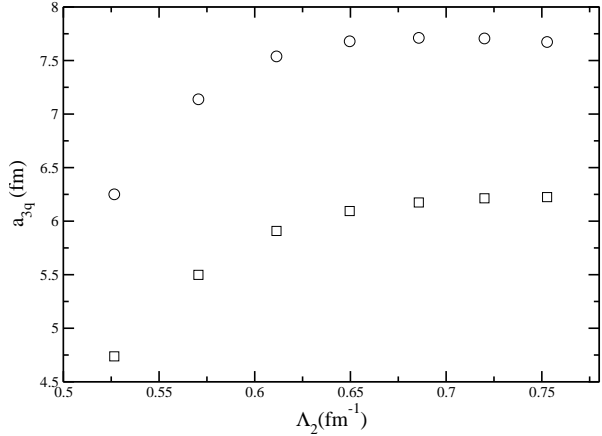}
\end{center} 
\caption{Scattering length $a_{3q}$ extracted from the spectrum of the
 trapped three-nucleon system in the channel $T=1/2$, $L=0$, $S=3/2$
as function of the cutoff $\Lambda_2$, for $\omega=1$ MeV:
%$J^{\pi}=\frac{3}{2}^{+}$ 
LO (circles) and NLO (squares). }
\label{scat_hw_1}
\end{figure}

We now consider the procedure to extract $a_{3q}$ for different values 
of the two-body cutoff and HO frequency.
Extracted values of $a_{3q}$ at LO and NLO as a function of the two-body 
cutoff are shown 
in Fig. \ref{scat_hw_1} for $\omega=1$ MeV.
At a fixed  $\omega$, the scattering length $a_{3q}$ should converge as 
the two-body cutoff is increased (since the
energies of the three-nucleon system converge). 
For each $\omega$, we perform extrapolations to obtain the value 
$a_{3q}({\infty})$ 
which would correspond to $\Lambda_2 \rightarrow \infty$. 
We use for this purpose the trial function
\begin{eqnarray}
\frac{1}{a_{3q}}&=& \frac{1}{a_{3q}({\infty})}
+\frac{\alpha_1}{\Lambda_2^{p_1}}
+\frac{\alpha_2}{\Lambda_2^{p_2}}, 
\label{fit2} 
\end{eqnarray}
where $p_{1,2}$ and $\alpha_{1,2}$ are parameters,
which we fit to the six values of the scattering length obtained at the 
largest cutoffs.

%%%%%%%%%%%%%%%%%%%%%%%%%%%%%%%%%%%%%%%%%%%%%%%%%%%%%%%%%%%%%%
%%%%%%%%%%%%%%%%%%%%%%%%%%%%%%%%%%%%%%%%%%%%%%%%%%%%%%%%%%%%%%

\begin{figure}[t]
\begin{center}
%\vspace*{-3cm} 
\includegraphics[scale=0.33,angle=-90,clip=true]{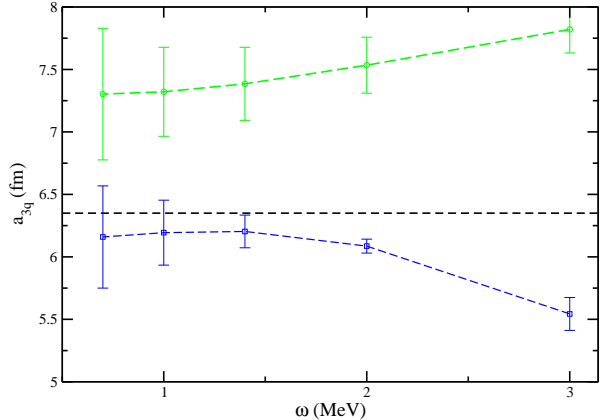}
\end{center} 
\caption{Extrapolated values $a_{3q}({\infty})$ of the 
quartet scattering length  
for different values of $\omega$: LO (circles) and NLO (squares).
The error bars correspond to the standard error
obtained using the software Gnuplot.
The horizontal dotted line marks the experimental value \cite{d_n_exp}.}
\label{scat_length}
\end{figure}

Results of the extrapolation 
can be seen in Fig. \ref{scat_length},
where $a_{3q}({\infty})$ is plotted as a function of $\omega$.
For HO frequencies from about 0.4 MeV to about 2 MeV
the scattering length is such that
$7.30~\mathrm{fm} \leq a_{3q}^{LO}(\infty) \leq 7.53 ~\mathrm{fm}$
and 
$6.08~\mathrm{fm} \leq a_{3q}^{NLO}(\infty) \leq 6.16~\mathrm{fm}$
for the trial function (\ref{fit2}).
Had we used a different trial function with fewer parameters,
such as for instance the function (\ref{fit2}) with
$\alpha_2=0$, we would have obtained 
$7.71~\mathrm{fm} \leq a_{3q}^{LO}(\infty) \leq 7.88 ~\mathrm{fm}$ 
and 
$6.20~\mathrm{fm} \leq a_{3q}^{NLO}(\infty) \leq 6.24~\mathrm{fm}$.
While for larger traps we are closer to the continuum limit,
our error in the scattering length increases.
First,
as $\omega$ gets smaller, the imprecision on the value of $a_{3q}$ stemming 
from the imprecision of the energy ({\it{i.e.}}, the difference between the 
values for finite $N_2^{max}$ and $N_2^{max}\to \infty$) are enhanced. 
This can be understood by noticing that in Eq. (\ref{eq:scat_3b}) 
the energy appears with 
$\omega$ in the denominator. 
Second,
numerical imprecision also arises
as $\omega$ gets smaller 
since the extrapolation to $a_{3q}({\infty})$ is performed in these cases  
from data at lower $\Lambda_2$.
Nevertheless we see a kind of plateau 
in the value of $a_{3q}({\infty})$ for small $\omega$.
With all imprecisions taken into account we can conclude that 
the results at NLO are in good agreement 
with the experimental value 
$a_{3q}=6.35 \pm 0.02 ~\rm{fm}$ \cite{d_n_exp} 
and with previous EFT calculations \cite{d_n_eft}. 
 
%%%%%%%%%%%%%%%%%%%%%%%%%%%%%%%%%%%%%%%%%%%%%%%%%%%%%%%%%%%%%%
%%%%%%%%%%%%%%%%%%%%%%%%%%%%%%%%%%%%%%%%%%%%%%%%%%%%%%%%%%%%%%

It might seem more natural to extract the scattering length 
directly from the extrapolated energies
$E_3(\infty)$ obtained in the fit (\ref{eq:fitE}).
After having tried this method, we concluded that it could not give 
meaningful results: the behaviour of $a_{3q}$ as a function of $\omega$ 
looked completely random. We believe that this is due to the fact
that the extrapolated energies are not precise enough, because
$a_{3q}$ is strongly dependent on the input energies.

\subsection{The channel $T=1/2$, $J^{\pi}=1/2^{+}$}
\label{sec_1_half}

We now consider the case of three nucleons with $T=1/2$
and orbital angular momentum and spins coupled
to total angular momentum $J^{\pi}=1/2^{+}$, the triton channel.
The calculation proceeds along the same lines discussed in detail
in the previous subsection, except for the role played by
the three-body force, which is similar
to that for three bosons \cite{hanstrap}.

\begin{figure}[t]
\begin{center}
\includegraphics[scale=0.4,angle=0,clip=true]{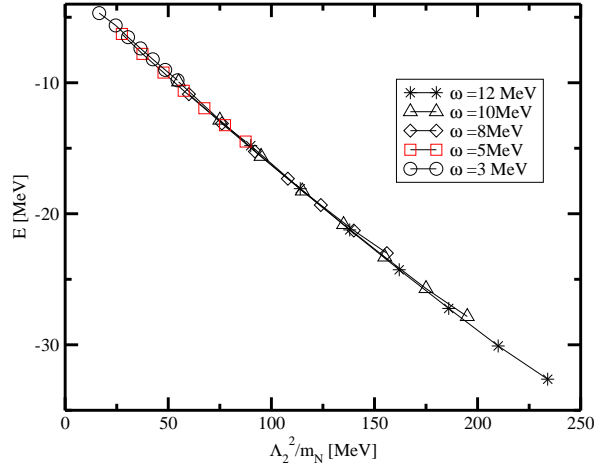}
\end{center} 
\caption{Ground-state energy of the trapped three-nucleon 
system with
$T=1/2$ and $J^{\pi}=1/2^{+}$
as a function of $\Lambda_2^2/m_N$,
for different frequencies $\omega$.  
Calculations are performed at LO but {\it without} a three-nucleon force.}
\label{triton_no_3b}
\end{figure}

The non-derivative three-nucleon potential (\ref{3BF}) contributes
in this channel and is known to be necessary for RG invariance at LO,
at least in free space \cite{pionlessT}.
Since renormalization concerns UV momenta, it is not expected to 
be affected by the trap.
We have confirmed this fact by examining the 
ground state of the three-nucleon system 
at various values of $\omega$ in a calculation at LO but {\it without}
a three-nucleon force. 
As before,
for a fixed two-body cutoff $\Lambda_2$
we increase the three-body model space until convergence is reached. 
Figure  \ref{triton_no_3b}
shows the ground-state energy as a function of $\Lambda_2^2/m_N$.
We can clearly see that as $\Lambda_2$
increases, the ground-state energy decreases roughly linearly. 
Results 
for different values of $\omega$ but the same two-body cutoff 
$\Lambda_2$ are close to each other, which
is a sign of the fact that the short-range two-nucleon interaction 
is much stronger than the long-range HO potential.
This  
illustrates the collapse of the three-nucleon system in this channel
when only a two-nucleon force is included in the pionless EFT 
\cite{pionlessT}.

%%%%%%%%%%%%%%%%%%%%%%%%%%%%%%%%%%%%%%%%%%%%%%%%%%%%%%%%%%%%%%
%%%%%%%%%%%%%%%%%%%%%%%%%%%%%%%%%%%%%%%%%%%%%%%%%%%%%%%%%%%%%%

\begin{figure}[tb]
\begin{minipage}[t]{75mm}
\centerline{\includegraphics[scale=0.38,clip=true]{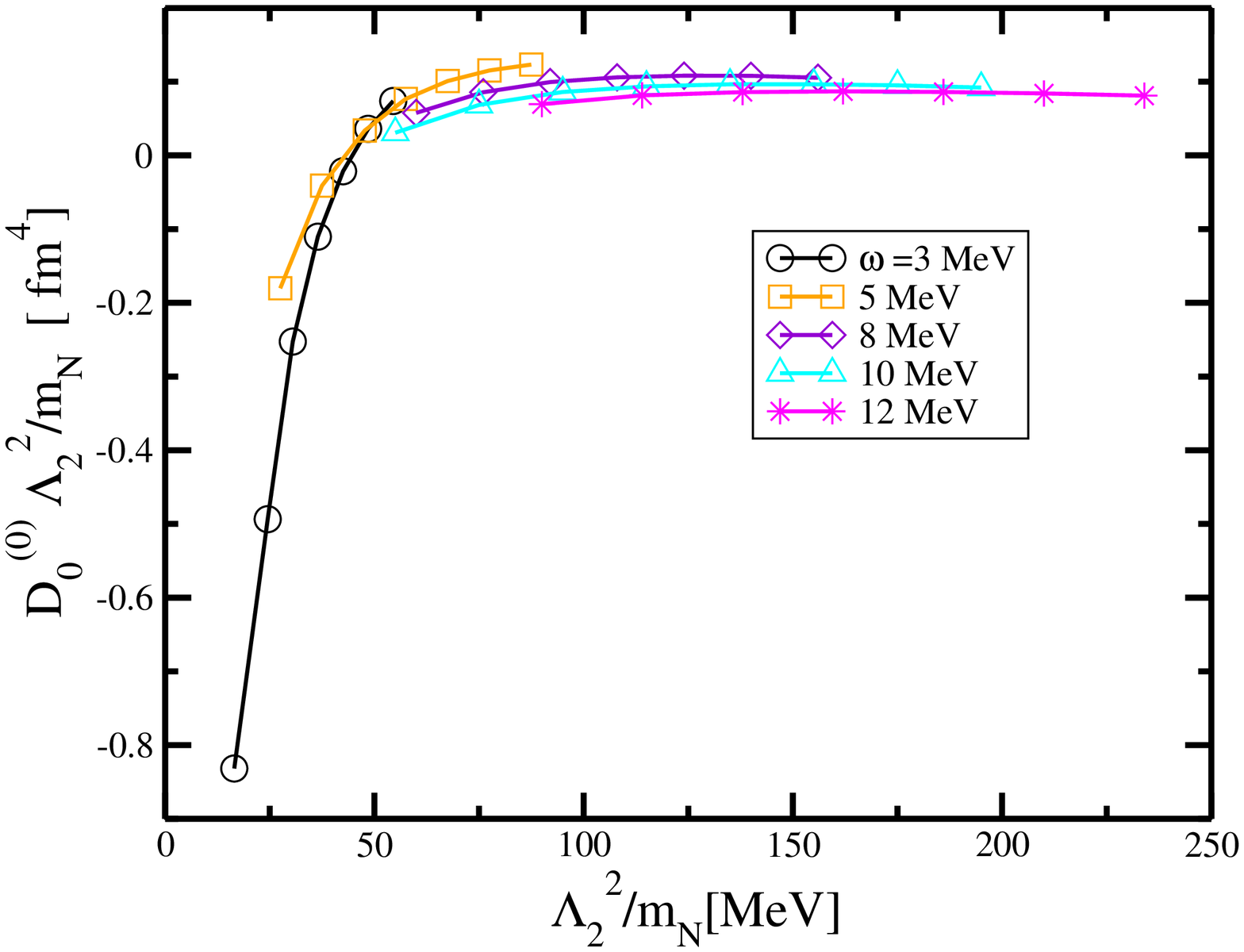}}
\end{minipage}
\hspace{\fill}
\begin{minipage}[t]{80mm}
\centerline{\includegraphics[scale=0.38,clip=true]{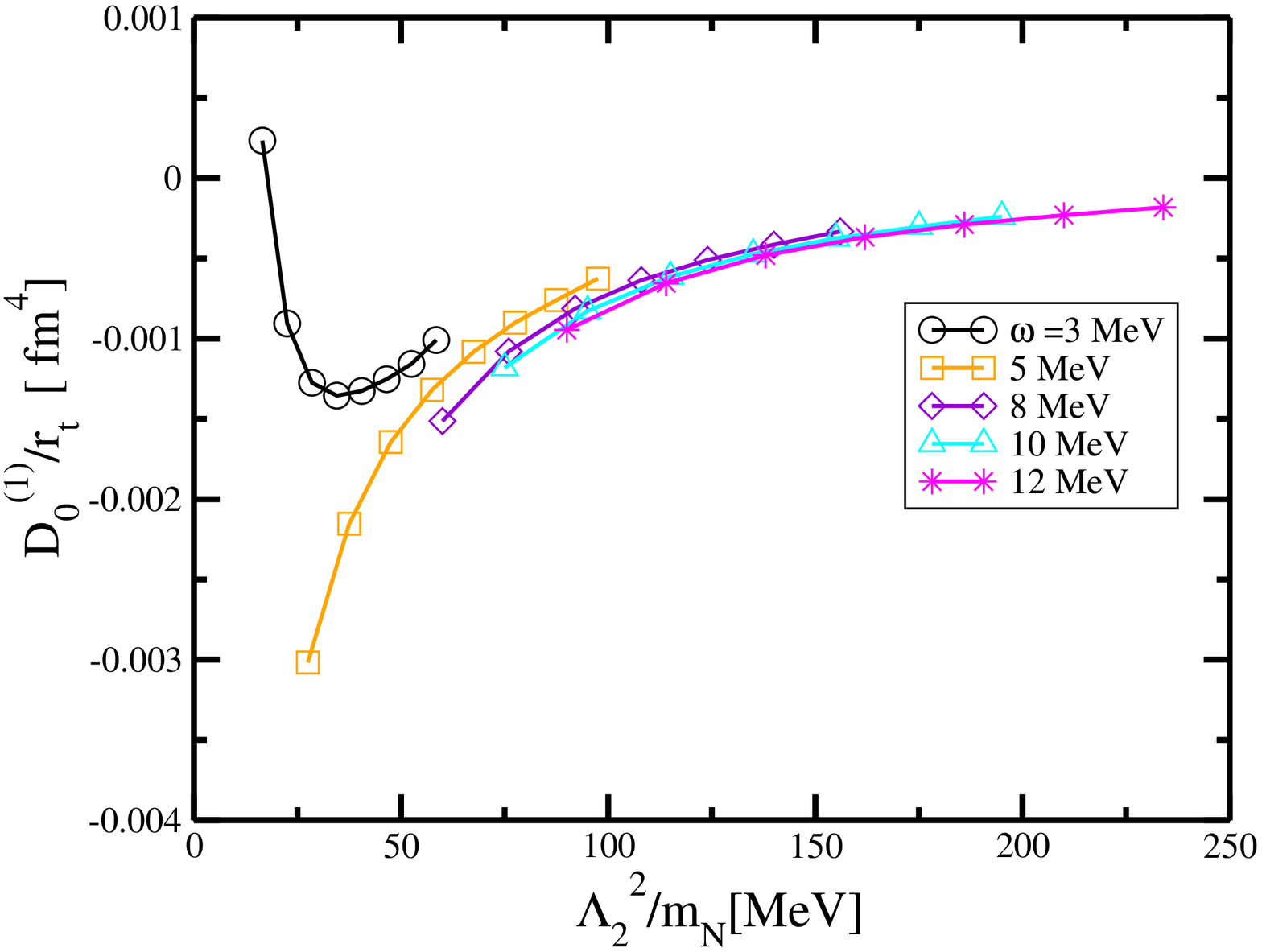}}
\end{minipage}
\caption{
Three-nucleon coupling constants as function of $\Lambda_2^2/m_N$, 
for different values of $\omega$:
$D_0^{(0)}\Lambda_2^2/m_N$ (left panel) and 
$D_0^{(1)}/r_t$ (right panel).
}
\label{D0}
\end{figure}

Such a dramatic cutoff dependence shows that short-distance
physics has not been accounted for properly. 
Renormalization can be achieved by 
including the non-derivative three-nucleon potential in Eq. (\ref{3BF})
already at LO.
We choose to determine the coefficient $D_0$ 
so that the lowest energy 
of the three nucleons in the trap 
is fixed at the experimental value of 
the triton binding energy,
$E_{t}=-8.482$ MeV. 
It is convenient to
introduce a cutoff $N_{3}^{cut}$
above which the three-nucleon force is switched off. 
This means that 
the three-body force does not play a role for 
configurations with $N_3>N_3^{cut}$. Nevertheless, $D_0$
depends on both $N_{3}^{cut}$ and $N_3^{max}$ just as the energy obtained  
with only the two-nucleon force depends on $N_3^{max}$
before convergence is reached.  
We take $N_{3}^{cut}=N_{2}^{max}$,
and for each $N_{2}^{max}$, $D_{0}$ is adjusted to 
the triton binding energy. 
For large enough $N_3^{max}$,
the three-body force
becomes independent on $N_{3}^{max}$,
\begin{equation}
D_0(N_{3}^{cut},N_{3}^{max})\rightarrow D_0(N_{3}^{cut}).
\end{equation}
We again split the running of $D_0(N_{2}^{max})$ 
into the various orders,
$D_0(\Lambda_2)=D_0^{(0)}(\Lambda_2)+D_0^{(1)}(\Lambda_2)+\ldots$
The running of $D_0^{(0)}(\Lambda_2)$ and $D_0^{(1)}(\Lambda_2)$
are shown in Fig. \ref{D0} for different values of $\omega$.
The LO three-nucleon force becomes repulsive for  
$\Lambda_2 \simge 220$ MeV.
We expect to see a limit cycle \cite{pionlessT} in the 
behaviour of the coupling $D_0^{(0)}\Lambda_2^2$ as a 
function of $\Lambda_2$.
However, the maximum cutoff we were able to consider here 
($\Lambda_2^2/m_N\sim 230$ MeV) is only approximately half 
the value
where the second branch of the limit cycle appears 
\cite{pionlessT}.

%%%%%%%%%%%%%%%%%%%%%%%%%%%%%%%%%%%%%%%%%%%%%%%%%%%%%%%%%% 
%%%%%%%%%%%%%%%%%%%%%%%%%%%%%%%%%%%%%%%%%%%%%%%%%%%%%%%%%%%%%%

\begin{figure}[tb]
\begin{minipage}[t]{75mm}
\centerline{\includegraphics[scale=0.33,angle=-90,clip=true]{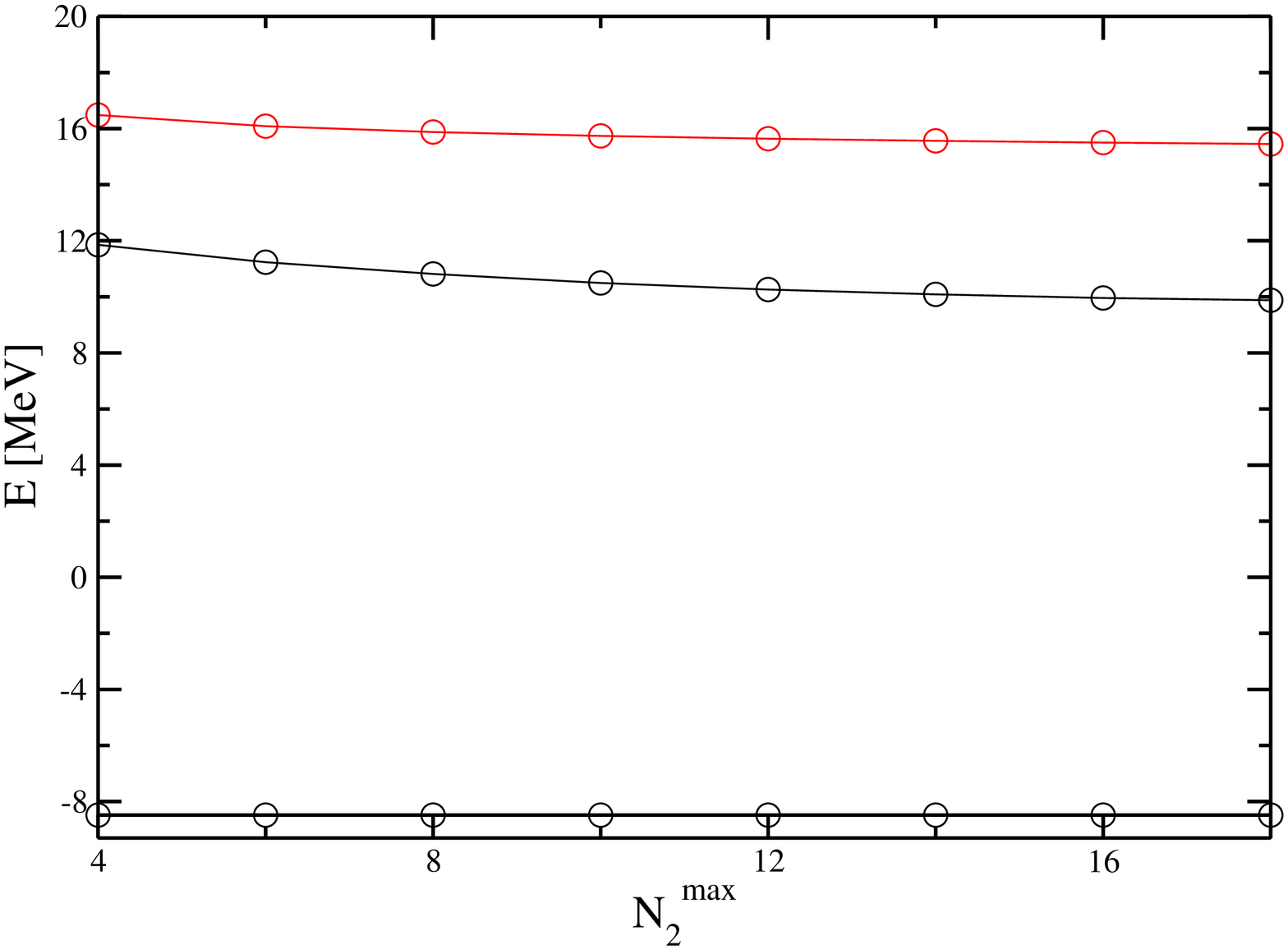}}
\end{minipage}
\hspace{\fill}
\begin{minipage}[t]{80mm}
\centerline{\includegraphics[scale=0.33,angle=-90,clip=true]{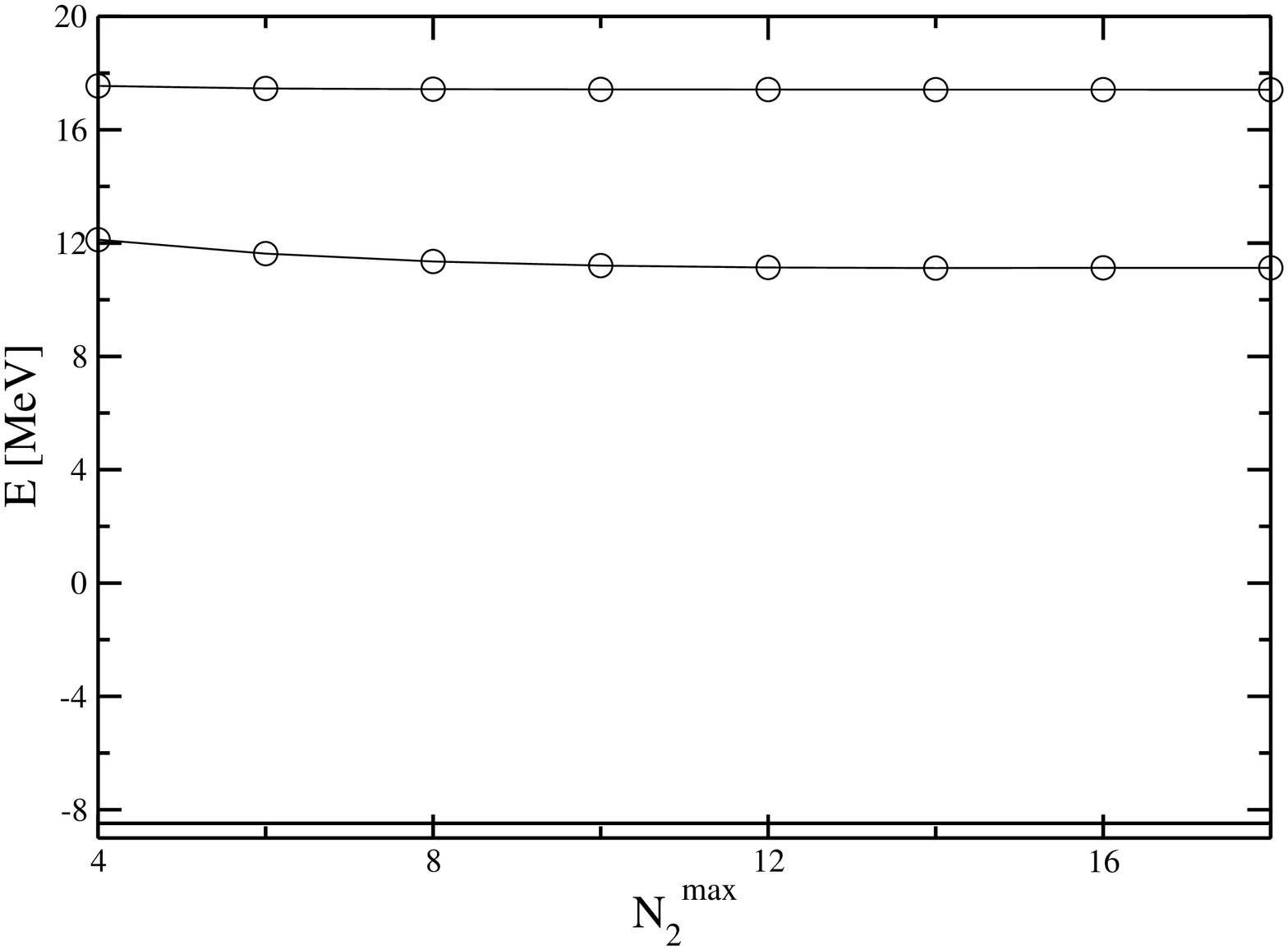}}
\end{minipage}
\caption{Energies of the ground state and first two excited states 
for the three-nucleon system coupled to 
$T=1/2$, $J^{\pi}=1/2^{+}$
as a function of  the two-body model-space size 
$N_2^{max}$, for $\omega=5$ MeV: LO (left panel) and NLO (right panel).
The three-body force is adjusted such that the ground state is fixed at 
the experimental value of the triton binding energy \cite{tritonBE}.
}
\label{exc_triton}
\end{figure}

Figure \ref{exc_triton} shows the LO and NLO
energies of the first three states 
in a trap with $\omega=5$ MeV.
While the ground state is fixed at the experimental value of
the triton binding energy, the other states converge to 
positive energy values, as befits continuum states in free space.
Again, results are similar for other HO frequencies.

%%%%%%%%%%%%%%%%%%%%%%%%%%%%%%%%%%%%%%%%%%%%%%%%%%%%%%%%%% 
%%%%%%%%%%%%%%%%%%%%%%%%%%%%%%%%%%%%%%%%%%%%%%%%%%%%%%%%%%%%%%

\begin{figure}[t]
\begin{center}
\includegraphics[scale=0.33,angle=-90,clip=true]{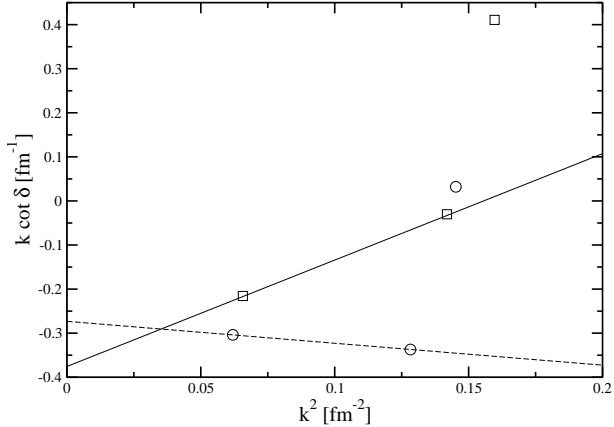}
\end{center} 
\caption{$k \cot \delta_3(k)$ in the $L=0$, $S=1/2$ channel 
obtained from the lowest energies at LO (circles) and NLO (squares)
as a function of $k^2$, for $\omega=1$ MeV and $N_2^{max}=18$.  
The dashed (full) line corresponds to a LO (NLO) linear fit 
to the two lowest energies.
} 
\label{doubletphases}
\end{figure}

{}From the scattering states we can again attempt to extract
the $S$-wave $nd$ phase shifts using Eq. (\ref{eq:scat_3b}).
The result is shown in Fig. \ref{doubletphases} for $\omega=1$ MeV 
and $N_2^{max}=18$.
By fitting the $k \cot \delta_3(k)$ with a first-degree 
polynomial in $k^2$, as in Sec. \ref{sec_3_half}, 
we can extract the doublet scattering length $a_{3d}$
using the two lowest energies. 
We then obtain $a_{3d}^{LO}=3.66$ fm and 
$a_{3d}^{NLO}=2.66$ fm.
This is far larger than the experimental value 
$a_{3d}=0.65\pm0.04$ fm \cite{d_n_exp} 
and results obtained with pionless EFT 
in the continuum \cite{pionlessT}. 
On the other hand,
using the second and third values of the 
phase shift gives much smaller values for the scattering length,
$a_{3d}^{LO}=0.319$ fm and $a_{3d}^{NLO}=0.281$ fm. 
Contrary to the quartet channel,
the values for $a_{3d}$ depend {\it strongly}
on which energies they are extracted from.
Possibly the energies considered here
are not small enough to prevent higher-order ERE terms  
from spoiling the extraction of $a_{3d}$. 
This would explain why 
the third values of the phase shift in Fig. \ref{doubletphases}
is far away, at both LO and NLO, from the fit of $k \cot \delta_3(k)$ 
obtained from the two lowest energies. 
A solution to overcome this problem would be to 
use larger
computer resources to perform calculations at very low values 
of $\omega$ but large values of $N_{2}^{max}$, and thus 
obtain better converged results at small enough energies.

%%%%%%%%%%%%%%%%%%%%%%%%%%%%%%%%%%%%%%%%%%%%%%%%%%%%%%%%%%%%%%
%%%%%%%%%%%%%%%%%%%%%%%%%%%%%%%%%%%%%%%%%%%%%%%%%%%%%%%%%%%%%%

\section{Conclusions and Outlook}
\label{sec_last}

We have presented an extension to nucleons of the work 
in Refs. \cite{trap0,us1,us2,us3,hanstrap}, in which the inter-particle 
interactions in a harmonic-oscillator basis are constructed within 
an EFT framework, by trapping the system in
a HO potential. 
This approach is designed to improve upon the work of Ref. \cite{LO_eft_nuc}
by using a procedure to fix the two-body parameters from the two-body data, 
and thus more 
efficiently going beyond leading order. 
We considered explicitly here interactions up to NLO.

We have illustrated in the two-nucleon system the renormalization procedure 
developed in Refs. \cite{trap0,us1,us2,us3,hanstrap}. We noted that
the scattering properties can be recovered from the discrete spectra, 
as long as the trap length parameter $b$ is much larger than the range 
of the interaction. As in the continuum, we were able to demonstrate 
systematic improvement of observables order by order.

In the three-nucleon system, we showed the extent to which
scattering information can be recovered from the discrete levels
of the trap.
We have presented results for the 
$T=1/2$, $J^{\pi}=3/2^{+}$  
channel, where we have shown that, as in free space, no three-nucleon 
interaction is needed to renormalize the system
at LO and NLO. We have estimated 
the quartet scattering length 
for nucleon-deuteron scattering.
Results at NLO are in good agreement with 
the experimental value and
previous EFT calculations. 
We also showed the collapse of the system  in the $T=1/2$, $J^{\pi}=1/2^{+}$ 
channel when no three-nucleon force is included. 
This work opens the door for further development in describing
scattering processes from bound-state physics, providing
an alternative to other methods under development \cite{scat}.

For the future, we plan to extend this work to 
$^4$He and $^6$Li, 
in order to test whether the reasonable agreement with experiment found 
in Ref. \cite{LO_eft_nuc} was accidental or can be improved at NLO, 
thus testing the limits of the pionless EFT with increasing the number 
of nucleons. We also intend to apply 
our method to the EFT in which the pion degrees of freedom are 
introduced explicitly, which should increase the reach of nuclear EFT.

\acknowledgments
UvK thanks Hans Hammer for useful discussions.
We are thankful to South Africa's
National Institute for Theoretical Physics (NITheP)
in Stellenbosch for hospitality during part of this work.
This research was supported by
the US NSF under grant PHY-0854912 (B.R.B. and J.R.)
and by the US DOE under grants DE-FG02-97ER41014 (I.S.), 
DE-FC02-07ER41457 (I.S.), and DE-FG02-04ER41338 (J.R. and U.v.K.).


\begin{thebibliography}{99}

\bibitem{NCSM} 
P. Navr\'atil, J.P. Vary, and B.R. Barrett, 
Phys. Rev. Lett. {\bf 84} (2000) 5728; 
Phys. Rev. C {\bf 62} (2000) 054311; 
P. Navr\'atil, S. Quaglioni, I. Stetcu, and B.R. Barrett, 
J. Phys. G \textbf{36} (2009) 083101.

\bibitem{LO_eft_nuc}
I. Stetcu, B.R. Barrett, and U. van Kolck, 
Phys. Lett. B \textbf{653} (2007) 358.

\bibitem{us3}
I. Stetcu, J. Rotureau, B.R. Barrett, and U. van Kolck, 
J. Phys. G {\bf 37} (2010) 064033.

\bibitem{ARNPSreview} 
P.F. Bedaque and U. van Kolck, 
Ann. Rev. Nucl. Part. Sci. {\bf 52} (2002) 339.

\bibitem{aleph}
U. van Kolck, 
hep-ph/9711222,
in A. Bernstein, D. Drechsel, and T. Walcher (eds.),
Proceedings of the Workshop on Chiral Dynamics 1997, Theory and Experiment, 
Springer-Verlag, Berlin, 1998;
Nucl. Phys. A {\bf 645} (1999) 273; 
D.B. Kaplan, M.J. Savage, and M.B. Wise, 
Phys. Lett. B {\bf 424} (1998) 390; 
J. Gegelia,
nucl-th/9802038;
J.-W. Chen, G. Rupak, and M.J. Savage,
Nucl. Phys. A {\bf 653} (1999) 386; 
D.R. Phillips, G. Rupak, and M.J. Savage,
Phys. Lett. B {\bf 473} (2000) 209. 

\bibitem{d_n_eft}
P.F. Bedaque and U. van Kolck,
Phys. Lett. B {\bf 428} (1998) 221;
P.F. Bedaque, H.-W. Hammer, and U. van Kolck, 
Phys. Rev. C {\bf 58} (1998) R641;
F. Gabbiani, P.F. Bedaque, and H.W. Grie{\ss}hammer,
Nucl. Phys. A {\bf 675} (2000) 601;
M.C. Birse, 
nucl-th/0509031.

\bibitem{pionlessT}
P.F. Bedaque, H.-W. Hammer, and U. van Kolck, 
Nucl. Phys. A {\bf 676} (2000) 357;
H.-W. Hammer and T. Mehen,
Phys. Lett. B {\bf 516} (2001) 353;
P.F. Bedaque, G. Rupak, H.W. Grie{\ss}hammer, and H.-W. Hammer,
Nucl. Phys. A {\bf 714} (2003) 589;
H.W. Grie{\ss}hammer, 
Nucl. Phys. A {\bf 744} (2004) 192;
Nucl. Phys. A {\bf 760} (2005) 110;
Few-Body Syst. {\bf 38} (2006) 67;
I.R. Afnan and D.R. Phillips,
Phys. Rev. C {\bf 69} (2004) 034010;
T. Barford and M.C. Birse,
J. Phys. A {\bf 38} (2005) 697;
L. Platter, 
Phys. Rev. C {\bf 74} (2006) 037001.

\bibitem{pionless4}
L. Platter, H.-W. Hammer, and U.-G. Mei{\ss}ner, 
Phys. Lett. B {\bf 607} (2005) 254;
J. Kirscher, H.W. Grie{\ss}hammer, D. Shukla, and H.M. Hofmann, 
Eur. Phys. J. A {\bf 44} (2010) 239.

\bibitem{trap0}
I. Stetcu, B.R. Barrett, U. van Kolck, and J.P. Vary, 
Phys. Rev. A \textbf{76} (2007) 063613.

\bibitem{us1}
I. Stetcu, J. Rotureau, B.R. Barrett, and U. van Kolck, 
Ann. Phys. {\bf 325} (2010) 1644.

\bibitem{us2}
J. Rotureau, I. Stetcu, B.R. Barrett, M.C. Birse, and U. van Kolck,
Phys. Rev. A {\bf 82} (2010) 032711.

\bibitem{hanstrap}
S. T\"olle, H.-W. Hammer, and B.C. Metsch,
Comptes Rendus Physique {\bf 12} (2011) 59.

\bibitem{luu}
T. Luu, M.J. Savage, A. Schwenk, and J.P. Vary,  
Phys. Rev. C {\bf 82} (2010) 034003.

\bibitem{3bozos}
P.F. Bedaque, H.-W. Hammer, and U. van Kolck, 
Phys. Rev. Lett. {\bf 82} (1999) 463;
Nucl. Phys. A {\bf 646} (1999) 444;
R.F. Mohr, R.J. Furnstahl, R.J. Perry, K.G. Wilson, and H.-W. Hammer,
Ann. Phys. {\bf  321} (2006) 225;
L. Platter and D.R. Phillips, 
Few-Body Syst. {\bf 40} (2006) 35.

\bibitem{ERE}
H.A. Bethe,
Phys. Rev. \textbf{76} (1949) 021603. 

\bibitem{ERE_ref}
J.J. de Swart, C.P.F. Terheggen, and V.G.J. Stoks, 
nucl-th/9509032;
D.E. Gonzalez Trotter {\it et al.}, 
Phys. Rev. Lett. {\bf 83} (1999) 409.

\bibitem{HT}
T. Busch, B.-G. Englert, K. Rz\c{a}\.{z}ewski, and M. Wilkens,
Found. Phys. {\bf 28} (1998) 549;
S. Jonsell, 
Few-Body Syst. {\bf 31} (2002) 255;
D. Blume and C.H. Greene,
Phys. Rev. A {\bf 65} (2002) 043613;
M. Block and M. Holthaus,
Phys. Rev. A {\bf 65} (2002) 052102;
E.L. Bolda, E. Tiesinga, and P.S. Julienne,
Phys. Rev. A {\bf 66} (2002) 013403;
A. Bhattacharyya and T. Papenbrock, 
Phys. Rev. A \textbf{74} (2006) R041602.

\bibitem{pseudo}
E. Fermi,
Ric. Scientifica {\bf 7} (1936) 13; 
G. Breit, 
Phys. Rev. \textbf{71} (1947) 215;
K. Huang and C.N. Yang,
Phys. Rev. \textbf{105} (1957) 767.
 
\bibitem{mehen}
T. Mehen,
Phys. Rev. A {\bf 78} (2008) 013614.

\bibitem{nijmegen}
V.G.J. Stoks, R.A.M. Klomp, M.C.M. Rentmeester, and J.J. de Swart, 
Phys. Rev. C {\bf 48} (1993) 792.

\bibitem{petr}
P. Navr\'atil, G.P. Kamuntavi\v{c}ius, and B.R. Barrett, 
Phys. Rev. C {\bf 61} (2000) 044001.

\bibitem{tritonBE}
J.E. Purcell, J.H. Kelley, E. Kwan, C.G. Sheu, and H.R. Weller,
Nucl. Phys. A {\bf 848} (2010) 1.

\bibitem{d_n_exp}
W. Dilg, L. Koester, and W. Nistler, 
Phys. Lett. B {\bf 36} (1971) 208.

\bibitem{scat}
K.M. Nollett, S.C. Pieper, R.B. Wiringa, J. Carlson, and G.M. Hale, 
Phys. Rev. Lett. \textbf{99} (2007) 022502;
S. Quaglioni and P. Navr\'atil, 
Phys. Rev. Lett. \textbf{101} (2008) 092501; 
Phys. Rev. C \textbf{79} (2009) 044606.  

\end{thebibliography}
\end{document}